\documentclass[11pt,preprint2]{aastex}
\usepackage{amsmath}

\newcommand{\myemail}{msp@phys.au.dk}

\shorttitle{Spin down of cooling hybrid stars}
\shortauthors{M. Stejner, F. Weber \& J. Madsen}

\begin{document}

\title{Signature of deconfinement with spin down compression in
cooling hybrid stars}

\author{Morten Stejner}
\affil{Department of Physics and Astronomy, University of Aarhus}
\affil{Ny Munkegade, Bld. 1520, DK-8000 Aarhus C, Denmark.}
\email{\myemail}
\and
\author{Fridolin Weber}
\affil{Department of Physics, San Diego State University}
\affil{5500 Campanile Dr, San Diego CA 92182-1233}
\and
\author{Jes Madsen}
\affil{Department of Physics and Astronomy, University of Aarhus}
\affil{Ny Munkegade, Bld. 1520, DK-8000 Aarhus C, Denmark.}


\begin{abstract}
\noindent The thermal evolution of neutron stars is coupled to their spin down
and the resulting changes in structure and chemical composition. This
coupling correlates stellar surface temperatures with rotational state
as well as time. We report an extensive investigation of the coupling
between spin down and cooling for hybrid stars which undergo a phase
transition to deconfined quark matter at the high densities present in
stars at low rotation frequencies. The thermal balance of neutron
stars is re-analyzed to incorporate phase transitions and the related
latent heat self-consistently, and numerical calculations are
undertaken to simultaneously evolve the stellar structure and
temperature distribution. We find that the changes in stellar
structure and chemical composition with the introduction of a pure
quark matter phase in the core delay the cooling and produce a period
of increasing surface temperature for strongly superfluid stars of
strong and intermediate magnetic field strength. The latent heat of
deconfinement is found to reinforce this signature if quark matter is
superfluid and it can dominate the thermal balance during the
formation of a pure quark matter core. At other times it is less
important and does not significantly change the thermal evolution.
\end{abstract}



\section{Introduction}
The chemical composition of neutron stars at densities beyond nuclear
saturation remains uncertain with alternatives ranging from purely
nucleonic compositions through hyperon or meson condensates to
deconfined quark matter -- see e.g. \cite{Weber:2005} and
\cite{Page:2006} for recent reviews with emphasis on quark matter. A
future understanding of neutron star structure gained through
confrontation of theoretical models with the now steadily growing body
of observational facts will therefore simultaneously constrain
fundamental elements of particle and nuclear physics
\citep{Lattimer:2007}. The interlinked processes of spin down
and thermal cooling present intriguing prospects of gaining insight in
the properties of matter in neutron star cores by confrontation with
soft X-ray observations of thermal radiation from neutron star
surfaces as they both depend sensitively on and to some extent
determine the chemical composition.

As neutron stars spin down and contract, their structure and
composition change with the increasing density -- drastically if phase
boundaries are crossed so new forms of matter become possible. We
shall here investigate how this influences the thermal evolution of
hybrid stars which contain large amounts of deconfined quark
matter. The increasing density and changing chemical composition
further imply additional entropy production in bulk and the release of
latent heat as particles cross any phase boundaries present. We
therefore re-analyze the thermal equilibrium of compact stars to show
how mixed phases may be incorporated. We thus arrive at a natural
description of the latent heat of phase transitions in compact stars,
but also find -- through direct numerical calculations -- that unless
the stellar structure changes very rapidly the effects of latent heat
on the thermal evolution are insignificant when compared to those of
the changing chemical composition and the surface area reduction.\\

Neutron stars are extremely compact objects and densities in their
cores reach well in excess of the nuclear saturation density. At such
densities the distance between particles is on the order of the
characteristic range of the nuclear forces. Therefore, as was stressed
recently by \cite{Baym:2006}, perturbative treatments in terms of few-
or even many-body forces -- although highly successful in describing
the properties of matter below nuclear saturation -- are no longer
well defined in the core of neutron stars. Further, the relevant
degrees of freedom should include the appearance of hyperons and
possibly deconfined quarks, so treatments in terms of nucleons alone
must also be seen as approximative. Hyperons are expected to appear at
densities around $2-3\rho_0$, where $\rho_0 = 0.153$ fm$^{-3}$ is the
baryon density at nuclear saturation. The appearance of hyperons so
softens the equation of state that purely hadronic equations of state
may not allow stable models compatible with the accurately measured
masses of neutron stars in binary systems
\citep{Baldo:2003,Schulze:2006}. \cite{Schulze:2006} further
demonstrate that this conclusion is highly robust with respect to
different assumptions about hyperon interactions and the nucleonic
equation of state. At best these studies are strong arguments against
purely hadronic compositions and they certainly do show the relevance
of considering alternatives such as a
transition at high densities to deconfined quark matter.\\

Quark matter represents an entirely new type of matter -- as opposed
to just an additional degree of freedom as in the case of hyperons in
the hadronic phase -- and it cannot be assumed to soften the equation
of state to the same extent
\citep{Alford:2007a,Schaffner-Bielich:2007}. Hybrid stars can be
be consistent with the high masses and radii indicated by recent
observations (e.g. \cite{Ozel:2006},
\cite{Freire:2008,Freire:2008a,Freire:2007a}), and the quark matter
equation of state can fulfill the
constraints imposed by heavy-ion collision transverse flow data and
$K^+$ production (which nucleonic equations of state do not). Further,
\cite{Drago:2008} found that only stars with a quark matter component
can rotate stably without losing angular momentum by emission of
gravitational waves through the r-mode instability at the 1122 Hz
indicated by recent observations for the X-ray transient XTE J1739-285
\citep{Kaaret:2007}. This conclusion is tentative as the observed
neutron star spin frequency awaits confirmation, but it again shows
how the composition of neutron stars is an open question to be
determined by a confrontation of theory and observation, and that
a deconfined quark matter phase remains a viable alternative.\\

For definiteness we shall work with the equation of state suggested
by \cite{Glendenning:1992}; hereafter the G$^{300}_{180}$ equation of
state. While this equation of state is not sophisticated in its
treatment of the quark matter phase, it is illustrative in that it
allows a very large pure quark matter core with a transition through a
mixed phase at relatively low densities around 2-5$\rho_0$. For our
purposes it is interesting in that it implies drastic changes in
composition with spin down -- the pure quark matter core disappears at
high spin frequencies for instance -- and it therefore represents
something close to a limiting case. We shall then be able to
investigate both the effects of steady conversion of hadronic matter
to quark matter and the sudden appearance of a pure quark matter core
in the hadronic phase. 

In the so-called minimal
scenario, which excludes exotic and very rapid neutrino processes
\citep{Page:2004}, the internal temperature of neutron stars 
drops from beyond $10^{10}$~K to around $10^9$~K within a few minutes
after their birth, and neutrino cooling continues to dominate for at
least the next few thousand years until the internal temperature has
dropped below $10^8$~K and photon cooling takes over (see,
e.g. \cite{Page:2004, Yakovlev:2004a} for reviews). If the highly
efficient direct Urca neutrino emission (i.e., essentially beta decays,
$n\rightarrow p+e^-+\bar{\nu}_e$ and related processes in quark
matter) is active, the stars may cool very rapidly and reach very low
temperatures on a timescale of a few hundred years. As we shall see
the nuclear direct Urca process is active in the mixed phase of hybrid
stars, and may thus control the thermal evolution. These processes may
be suppressed by pairing of the participating particles however, and
further the extent of the mixed phase depends strongly on the rotation
frequency thus giving rise to a diverse range of possible cooling
paths.

The latent heat of the phase transition and the related release of
entropy in bulk with changing density are generally found to be of
smaller importance than the changing structure and chemical
composition. It does not significantly delay or enhance the cooling,
although it does briefly balance or even dominate the cooling terms
when the pure quark matter core is first formed for certain choices of
stellar parameters. The latent heat of the quark matter phase
transition was previously considered by \cite{Miao:2007, Miao:2007a} and,
\cite{Xiaoping:2008}. These authors took a different approach to calculate
the latent heat than what is discussed below and found very
significant heating terms which we do not recover.

The work of \cite{Reisenegger:1995} and lately \cite{Fernandez:2005}
considered the related process of roto-chemical heating for neutron
stars in which weak reactions are driven out of equilibrium by the
changing density. The resulting release of energy was found able to
maintain old stars at relatively high temperatures determined by their
rate of spin down. This term naturally appears in our equations for
the thermal balance and should be included in a full model. It is to
some extent complementary to the effects discussed here, but the
treatment of weak reactions beyond equilibrium is beyond our scope in
this work, and we shall assume chemical equilibrium throughout. We
return briefly to discuss this issue in Sect.~\ref{discuss}. \\

The link between the thermal evolution of neutron stars and
their spin down has become a subject of some interest as an
observational correlation between the temperatures and inferred
magnetic fields of neutron stars was recently discovered by
\cite{Pons:2007}. This was interpreted as evidence for magnetic field
decay and detailed work by \cite{Aguilera:2007, Aguilera:2008}
strongly supports this conclusion. An alternative interpretation was
offered recently by \cite{Niebergal:2007} in terms of magnetic flux
expulsion from color-flavor locked quark stars (a hypothetical class
of stars consisting entirely of quark matter which in this case is
assumed to be absolutely stable). The previously mentioned work of
\cite{Drago:2008} also indicates a link between the spin down and
cooling of hybrid stars. These correlations -- if they can indeed be
shown to exist -- complement the traditional cooling calculations
which relate only temperature and age, and they may help break the
degeneracy seen between such calculations with different underlying
assumptions about
the state of matter at very high density.\\

In the following, we shall first revisit the equations of
thermal balance for compact stars in Sect.~\ref{thequilibrium} to show
the effects of a time-dependent density and discuss how the presence
of phase transitions may be included. In Sect.~\ref{eosandmodel}, we
discuss the G$_{180}^{300}$ equation of state, the resulting stellar
models and some simple estimates for the additional terms in the
thermal balance equations in more detail. In
Sect.~\ref{calculations} we show the results of including these terms
in spherical isothermal cooling calculations. We conclude with a
discussion in Sect.~\ref{discuss}.

\section{Thermal Equilibrium in the Mixed Phase}\label{thequilibrium}
The thermal evolution of compact stars is determined by the equations
of local energy conservation and transport in the framework of general
relativity. These equations balance any energy radiated away from the
star by photons and neutrinos against changes in rest mass due to
nuclear reactions and changes in gravitational or internal energy as
the stellar structure evolves. In the most widely studied scenario the
stellar structure is assumed constant and the only energy source
available to neutron stars is then their original endowment of thermal
energy (e.g.,
\cite{van-Riper:1991,Schaab:1996,Page:2004,Page:2006a,Yakovlev:2004a}
and references therein). Neutrino production in the core and photon
emission from the stellar surface then ensures a monotonical and
sometimes very rapid cooling of the star depending strongly on any
superfluid properties of the core. But in addition to this a number of
powerful energy sources may play a role in delaying or even reversing
the cooling at various stages in a neutron stars life -- see
e.g. \cite{Schaab:1999} for an overview of possible sources and their
effects. Here we discuss the effects of including a phase transition
in the energy balance, which, as the stellar structure changes, must
then also include redistribution of entropy between the two phases as
their proportion changes as well as changes in surface and Coulomb
energy for transitions through a mixed phase.

Following \cite{Thorne:1966} or the equivalent discussion in
\cite{Weber:1999} we consider energy conservation for a spherical
shell inside of which is $a$ baryons and which itself contains $\delta
a$ baryons. The treatment given in this section therefore assumes
spherical symmetry which is sufficient to show the effects of a phase
transition on local energy conservation -- we shall return later to
discuss possible consequences of a multidimensional cooling
calculation. The shell under consideration will change its
internal energy during a coordinate time interval d$t$ by
\begin{eqnarray}\label{balance}
&&d(\textrm{internal energy})=\nonumber\\
&&(\textrm{amount of rest mass-energy converted
  to} \nonumber\\ 
&&\textrm{internal energy by reactions})\nonumber\\
+&&(\textrm{work done on shell by
gravitational}\nonumber\\ 
&&\textrm{forces to change its volume during} \nonumber\\ 
&&\textrm{quasi-static contraction})\nonumber\\ 
-&&(\textrm{energy radiated, conducted} \nonumber\\ 
&&\textrm{or convected away from the shell}).
\end{eqnarray}

It is important to note at this point that if the two phases are in
equilibrium -- as should certainly be expected for transitions
governed by strong reactions such as that from hadronic to quark
matter -- there is {\it no binding energy} involved in the transition
and therefore no contribution to the first term on the right hand side
of Eq.~(\ref{balance}). If this was the case and the phase transition did
involve a binding energy the two phases could not be in
equilibrium and the star would adjust itself accordingly -- eventually
becoming a strange star in the case of the quark matter transition if
quark matter was assumed bound relative to hadronic matter at zero
external pressure. In hybrid stars this is not the assumption however,
and so any latent heat evolved or absorbed in the phase transition
follows from the different thermodynamical properties of the
two phases as with any other phase transition.

In the mixed phase of the G$^{300}_{180}$ equation of state regions
containing negatively charged quark matter appear at densities of
about 2 times nuclear saturation and dominate completely at 5
times nuclear saturation. They allow the hadronic matter to
lower its isospin asymmetry energy and become positively charged by
including more protons with charge neutrality achieved globally. The
geometry and structure of the mixed phase is determined by a balance
between surface tension and Coulomb repulsion between regions of like
charge. For details on the phase transition we refer to
e.g. \cite{Glendenning:1992,Heiselberg:1993,Glendenning:2000,Glendenning:2001a,
Voskresensky:2003,Endo:2006}. 
   
Returning to our spherical shell of baryon number $\delta a$ we will
assume that its volume $V$ is large enough to contain a macroscopic number of unit
cells each containing a region filled by the rare phase whose presence
may then be considered a microscopic property of the equation of
state. The work done to change the shells volume during
contraction or expansion of the star must then include changes in
surface and Coulomb energy as well as the usual pressure term
\begin{equation}\label{work}
\mathrm{d}W=-P\mathrm{d}V + \alpha \mathrm{d}\mathbb{S}+\mathrm{d}E_\mathrm{C}
\end{equation} 
where $\alpha$ is the surface tension, $\mathbb{S}$ is the amount of
surface area in the shell dividing the two phases and $E_\mathrm{C}$
is the Coulomb energy contained in the shell. These terms can be
similarly included in the first law of thermodynamics which we write
in the same notation (units with $G=c=k_\mathrm{B}=1$ will be used
here and throughout this paper)
\begin{eqnarray}\label{firstlaw}
\mathrm{d}\epsilon=&\frac{P+\epsilon}{\rho}\mathrm{d}\rho+T\rho\mathrm{d}s
+\sum_k \mu_k \rho \mathrm{d}Y_k\nonumber\\ 
&+\rho\frac{\alpha
  \mathrm{d}\mathbb{S}}{\delta a}+\rho\frac{\mathrm{d}
  E_\mathrm{C}}{\delta a}
\end{eqnarray}   
where $s$ is the entropy per baryon and $Y_\mathrm{k}=\rho_k/\rho$ is
the fraction of the baryon number density in the form $k$ with $k$
running over all particle species and $\sum_k Y_k =1$. In the quark
matter phase each quark contributes by only $\frac{1}{3}$ to the
baryon number density and $s$ is then the entropy per three quarks. The
last two terms in Eq.~(\ref{firstlaw}) are the local densities of
surface and Coulomb energy written in a form useful for our purpose.

Inserting this in Eq.~(\ref{balance}) and using that the amount of energy
radiated, conducted or convected away from the shell can be written in
terms of the gradient of the total luminosity, $L_\mathrm{tot}$, we
show in Appendix~\ref{appendix} that local energy balance may be expressed as
\begin{equation}\label{balance2}
\frac{\mathrm{d}}{\mathrm{d}a} (L_\mathrm{tot} e^{2\Phi}) =
-e^{\Phi}\left [
  T\left(\frac{\mathrm{d}s}{\mathrm{d}t}\right)_{a}+\sum_k \mu_k \left(\frac{\mathrm{d}Y_k}{\mathrm{d}t}\right)_{a} \right ]
\end{equation} 
thus giving the contribution to $L_\mathrm{tot}$ from the shell in
terms of the entropy production and the difference between the
chemical potentials of particles participating in any reactions taking
place in the shell. $e^{2\Phi}$ is the time component of a spherically
symmetric metric 
\begin{equation}
ds^2=-e^{2\Phi}dt^2+e^{2\Lambda}dr^2+r^2d\theta^2+r^2\sin{\theta}^2d\Phi^2
\end{equation}
found from the general relativistic structure equations for compact
stars.

$L_\mathrm{tot}$ includes the neutrino
luminosity, but since neutrinos can be taken to immediately escape
from the star when they are created without converting into any other
form of energy along the way, they fulfill their own separate
equation of energy conservation
\begin{equation}\label{neutrinobalance}
\frac{\mathrm{d}}{\mathrm{d} a}(L_\nu e^{2\Phi}) = \frac{\epsilon_\nu}{\rho}e^{2\Phi}
\end{equation} 
where $L_\nu$ is the neutrino luminosity and $\epsilon_\nu$ is the neutrino emissivity; the rate per unit
volume at which neutrino energy is created. In neutron stars
convection is negligible compared to electron conduction and photon diffusion
and so the remainder of
$L_\mathrm{tot}$ can be shown to fulfill a transport equation
\citep{Thorne:1966,Weber:1999}  
\begin{equation}\label{transport}
\frac{\mathrm{d}}{\mathrm{d} a} \left(T e^\Phi \right) =
-\frac{3}{16\sigma}\frac{\kappa
  \epsilon}{T^3}\frac{Le^\Phi}{16\pi^2 r^4 \rho} 
\end{equation}
where $\sigma$ is the Stefan-Boltzmann constant, $\kappa$ is the
total thermal conductivity and $L=L_\mathrm{tot}-L_\nu$. At the
stellar center we must have $L_\mathrm{tot}(a=0)=0$ while at the
stellar surface $L$ must equal the total stellar photon luminosity
which may depend on assumptions about properties of the stellar
atmosphere or lack thereof.

Eqs.~(\ref{balance2})-(\ref{transport}) with the appropriate boundary
conditions can be solved to evolve the thermal structure of a stellar
model. They have exactly the same form as would be expected in the
absence of any phase transitions. However, the phase transition influences
the entropy density and this, as we shall see, gives rise to additional
terms in the heat balance including latent heat and the surface and
Coulomb energies. An equivalent form of Eq.~(\ref{balance2})
showing the contributions from surface and Coulomb terms more
explicitly can be found in Eq.~(\ref{cool1}) of Appendix~\ref{appendix}. Combining Eqs.~(\ref{balance2}) and
(\ref{neutrinobalance}), noting that $c_\mathrm{v}=\rho T(\partial
s/\partial T)_V$ and assuming constant structure and chemical
composition we also recover the standard cooling equation for static
stars 
\begin{equation}
\frac{\mathrm{d}}{\mathrm{d} a}(L e^{2\Phi})
  =-\rho^{-1}\left(\epsilon_\nu e^{2\Phi}+c_\mathrm{v}
  \frac{\mathrm{d}(Te^{\Phi})}{\mathrm{d} t}  \right)
\end{equation}

Eq.~(\ref{balance2}) is useful because it allows a
particularly simple analysis in the presence of a phase
transition. The first thing to note is, that particles crossing the
phase boundary would not contribute to the second term on the right
hand side of Eq.~(\ref{balance2}) if the two phases are in or close to
equilibrium, because their chemical potentials (or those of their
constituents) must then be continuous across the phase boundary. In
the following we therefore neglect this term, but we shall return to
discuss its potential importance for reactions not in equilibrium. The
first term is a different matter however. The entropy per particle is
a function of density, temperature and chemical composition which is
not required to be continuous across the phase boundary, so particles
making the transition will release or absorb heat accordingly. To see
how this works we write $s$ as a sum with bulk contributions from each
phase according to their volume or mass fraction as well as
contributions from the surface and Coulomb energies
\begin{align}
s&=\frac{1}{\rho}((1-\chi_\mathrm{v})S_{\mathrm{1}}+\chi_\mathrm{v}S_\mathrm{2}+S_\mathrm{S}+S_\mathrm{C})\\
&=(1-\chi_\rho)s_1+\chi_\rho s_2+\frac{1}{\rho}(S_\mathrm{S}+S_\mathrm{C})
\end{align}
where
$\rho=(1-\chi_\mathrm{v})\rho_\mathrm{1}+\chi_\mathrm{v}\rho_\mathrm{2}$
is the average of the two particle densities $\rho_{1,2}$ weighed by
the volume fraction of the dense phase, $\chi_\mathrm{v}$. $S_{1,2}$
is the bulk entropy density in each phase and $S_\mathrm{S}$ and
$S_\mathrm{C}$ are the surface and Coulomb contributions to the
entropy density respectively. We have further
introduced the baryon number fraction $\chi_\rho$ of baryons in the
dense phase in a sample of matter related to the volume fraction by
$\chi_\rho=\chi_\mathrm{v}(\rho_2/\rho)$, as well as the entropies
per baryon in the respective phases, $s_{1,2}=S_{1,2}/\rho_{1,2}$. At
constant $a$ and assuming the fraction of matter in the dense phase and
the particle densities do not depend on temperature we then have
\begin{eqnarray}\label{dsdt}
T\frac{\mathrm{d}s}{\mathrm{d}t}&=&\frac{c_\mathrm{v}}{\rho}\frac{\mathrm{d}T}{\mathrm{d}t}+T\frac{\mathrm{d}s}{\mathrm{d}\rho}\frac{\mathrm{d}\rho}{\mathrm{d}t}\nonumber\\&=&\frac{c_\mathrm{v}}{\rho}\frac{\mathrm{d}T}{\mathrm{d}t}+T\frac{\mathrm{d}\chi_\rho}{\mathrm{d}\rho}\frac{\mathrm{d}\rho}{\mathrm{d}t}\left(s_2-s_1\right)\nonumber\\
&&+(1-\chi_\rho)T\frac{\mathrm{d}\rho}{\mathrm{d}t}\frac{\mathrm{d}s_1}{\mathrm{d}\rho}
+\chi_\rho
T\frac{\mathrm{d}\rho}{\mathrm{d}t}\frac{\mathrm{d}s_2}{\mathrm{d}\rho}\nonumber\\
&&+T\frac{\mathrm{d}}{\mathrm{d}t}\left[\frac{1}{\rho}\left(S_\mathrm{S}+S_\mathrm{C}\right)\right]
\end{eqnarray}
where the heat capacity is again a weighed volume
average in the mixed phase 
\begin{equation}
c_\mathrm{v}=(1-\chi_\mathrm{v})c_\mathrm{V,1} +
\chi_\mathrm{v}c_\mathrm{V,2}\; .
\end{equation}

The latent heat absorbed by a particle crossing the boundary between
two phases in equilibrium is the temperature times the difference in
entropy per particle between the two phases, $q=T[s_2-s_1]$
\citep{Landau:1980}; we recover this in the second term on the right
hand side of Eq.~(\ref{dsdt}). If any term in Eq.~(\ref{dsdt}) is
negative, heat is evolved by this term which then heats the star and
adds to the luminosity $L$. In particular the latent heat is a heating
term for increasing density when the entropy per baryon of the dense
phase is less than that of the low density phase -- a situation which
will arise when considering superfluid quark matter.\\

For future reference we identify the terms in Eq.~(\ref{dsdt}) as
follows. $T\frac{\mathrm{d}\chi_\rho}{\mathrm{d}\rho}\frac{\mathrm{d}\rho}{\mathrm{d}t}(s_2-s_1)$
is identified as the latent
heat, $(1-\chi_\rho)T\frac{\mathrm{d}\rho}{\mathrm{d}t}\frac{\mathrm{d}s_1}{\mathrm{d}\rho}$
is identified as the hadronic bulk contribution, $\chi_\rho
T\frac{\mathrm{d}\rho}{\mathrm{d}t}\frac{\mathrm{d}s_2}{\mathrm{d}\rho}$
is identified as the quark bulk contribution, and the
last term $T(\mathrm{d/d}t)[(S_\mathrm{S}+S_\mathrm{C})/\rho]$ is identified as the surface and Coulomb contribution. Further we often
refer to these terms collectively as additional entropy production (or
release) beyond what would be expected at constant density.

We discuss in the following section how to calculate the bulk
entropy density of the hadronic and quark matter phases in the
relativistic mean field theory framework of the
$\mathrm{G}_{180}^{300}$ equation of state. For now let us just
remark that the entropy per particle for an ideal relativistic
degenerate Fermi gas is \citep{Landau:1980}
\begin{equation}\label{s}
s=\frac{(3\pi^2)^{\frac{2}{3}}}{3\hbar c}T\rho^{-\frac{1}{3}}=0.02\frac{T}{\mathrm{MeV}}\left(\frac{\rho}{\mathrm{fm}^{-3}}\right)^{-\frac{1}{3}}\;.
\end{equation}  
Taking the transition to quark matter as a transition between such
gases -- note that the bag constant does not contribute to the
entropy -- and remembering that each hadron contains three quarks
which further have a color degeneracy of three, the latent heat in
Eq.~(\ref{dsdt}) is of the order of 
\begin{align}\label{estimate}
10^{33}T_9^2\bigg[9\bigg(\frac{\rho_\mathrm{Q}}{\mathrm{fm}^{-3}}\bigg)^{-\frac{1}{3}}-&
\left(\frac{\rho_\mathrm{H}}{\mathrm{fm}^{-3}}\right)^{-\frac{1}{3}}\bigg]\\\nonumber
&\times\frac{\mathrm{d}a_\mathrm{Q}/\mathrm{d}t}{10^{57}/10^7\mbox{ yr}}\mbox{ erg s}^{-1}
\end{align}
where $T_9=T/10^9$~K, and the rate of baryons making the transition to
quark matter, $\mathrm{d}a_\mathrm{Q}/\mathrm{d}t$, was (arbitrarily)
scaled to an entire star being converted steadily over $10^7$
years. Unless the star is very hot or changing structure fast this is
a very modest contribution, and we further note that it is positive for
a contracting star and so acts to cool the star down. However the
sign is subject to the very rough assumption that both gases may be
treated as ideal Fermi gases for the purpose of calculating their
entropy, and it will change in a more detailed treatment. Specifically
the quark phase may be color-superconducting with energy gaps as high
as 100 MeV and corresponding critical temperatures of the order of
$10^{11}$~K. Below the critical temperature the quark specific heat
and entropy density would be exponentially suppressed and could be
ignored relative to the hadronic contribution in Eq.~(\ref{estimate})
which would then be a heating term. We shall return to both
possibilities in later sections.

We shall calculate the two bulk terms numerically in the following
sections, so for now we just note that from Eq.~(\ref{s}) the entropy
per baryon decreases with increasing density. In a contracting star
these terms will hence act as heating terms and locally be of the same
order of magnitude as the release or absorption of latent heat in
Eq.~(\ref{estimate}) -- but of course they contribute throughout the
star and are therefore potentially far more important than the latent
heat which is only significant in the mixed phase.

Since the surface and Coulomb energies are related by
$E_\mathrm{S}=2E_\mathrm{C}$ in equilibrium \citep{Glendenning:2000}, their contributions to the
thermodynamic potential, and hence to the entropy, are
similarly fixed in proportion, and we need only consider one of them
here. Specifically the surface part of the entropy may be found from
\citep{Landau:1980}
\begin{eqnarray}
&&\Omega=\Omega_0+\Omega_\mathrm{S}=\Omega_0+\alpha\mathbb{S}\\
&&S_\mathrm{S}=-\frac{\partial \Omega_\mathrm{S}}{\partial
  T}=-\mathbb{S}\frac{\partial \alpha}{\partial T}\;,
\end{eqnarray}
where $\Omega_\mathrm{S}$ is the surface contribution to the thermodynamic
potential $\Omega$. The surface
tension for the quark-hadron interface, $\alpha$, determines the
geometry and extent of the mixed phase \citep{Glendenning:1992,
 Heiselberg:1993, Voskresensky:2003, Endo:2006}. The surface tension
of strangelets in vacuum has been evaluated by \cite{Berger:1987}, but
the surface tension for the mixed phase remains essentially
unknown. It is commonly parameterized as being proportional to the
difference in energy density between the two phases and the length
scale $L\sim 1$ fm of the strong interaction \citep{Glendenning:2000} 
\begin{equation}
\alpha=K\times[\epsilon_\mathrm{Q}-\epsilon_\mathrm{H}]\times L\;.
\end{equation} 
Assuming for simplicity that $K$ and $L$ are constant we then get
\begin{align}
&\frac{\partial \alpha}{\partial T}=\alpha\frac{c_\mathrm{V,Q}-c_\mathrm{V,H}}{\epsilon_\mathrm{Q}-\epsilon_\mathrm{H}}\\
&\rho^{-1}(S_\mathrm{S}+S_\mathrm{C})=-\frac{3}{2}\frac{\epsilon_\mathrm{S}}{\rho}\frac{c_\mathrm{V,Q}-c_\mathrm{V,H}}{\epsilon_\mathrm{Q}-\epsilon_\mathrm{H}}
\end{align}
The corresponding term in Eq.~(\ref{dsdt}) is then of the order of the
surface energy per baryon times the ratio between thermal and total
energy density. It can therefore not be expected to contribute
significantly, and this expectation is confirmed by the numerical results.

\section{Equation of State and Stellar Models}\label{eosandmodel}
In our numerical work we have employed the rotating neutron star code
developed by Weber and the G$_{180}^{300}$ equation of state used by
\cite{Glendenning:1992,Glendenning:2000} and
\cite{Glendenning:2001a}. Here we shall briefly describe each of these
and the resulting stellar models.

The G$_{180}^{300}$ equation of state treats the deconfined quark
matter phase in a simple version of the bag model which ignores gluon
interactions. The confined hadronic phase is described in terms of the
mean field solution to a covariant Lagrangian that involves the baryon
octet interacting through scalar, vector and vector-isovector
mesons. The precise values of the coupling constants for the hadronic
part of the model correspond to the first set of Table 5.5 of
\cite{Glendenning:2000} or Table 1 of \cite{Glendenning:2001a}. For
details we refer to these works where the resulting equation of state as well as the underlying theory are
carefully described at zero temperature. Finite temperature expressions for pressure, energy and particle
densities can be found by reinserting the Fermi distribution in the
phase space integrals of their zero temperature expressions which we
shall not write explicitly here. We refer to
e.g. \cite{Glendenning:1990} for the full temperature dependent
expressions (this reference also includes gluon interactions to first
order which we ignore here). 

The entropy is calculated as
\begin{equation}
s=\frac{1}{\rho T}\left[P+\epsilon-\sum_i \mu_i\rho_i\right]
\end{equation}
using the finite temperature expressions outlined above and using the
highly accurate publicly available code described in
\cite{Miralles:1996} to solve the Fermi integrals. In keeping with our
assumption that temperatures remain too low to significantly influence the
chemical composition we neglect contributions from thermally exited
particle-antiparticle pairs. 

\begin{figure}[!htb]
\plotone{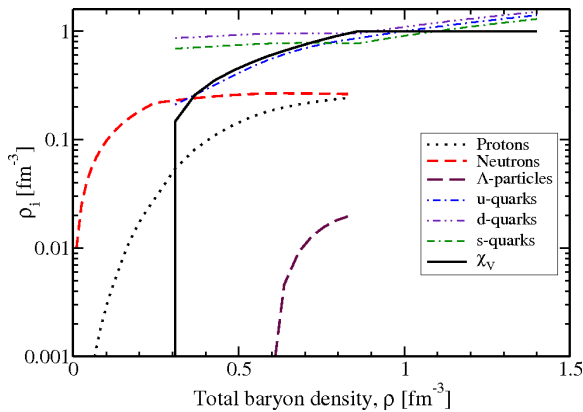}
\caption{Chemical composition of the G$^{300}_{180}$ equation of state
  as a function of total baryon density. Individual particle densities
  $\rho_i$ refer to the density in regions filled with the respective
  phase. $\chi_v$ is the (dimensionless) volume fraction of quark matter.}
\label{composition}
\end{figure}

\begin{figure}[!htb]
\plotone{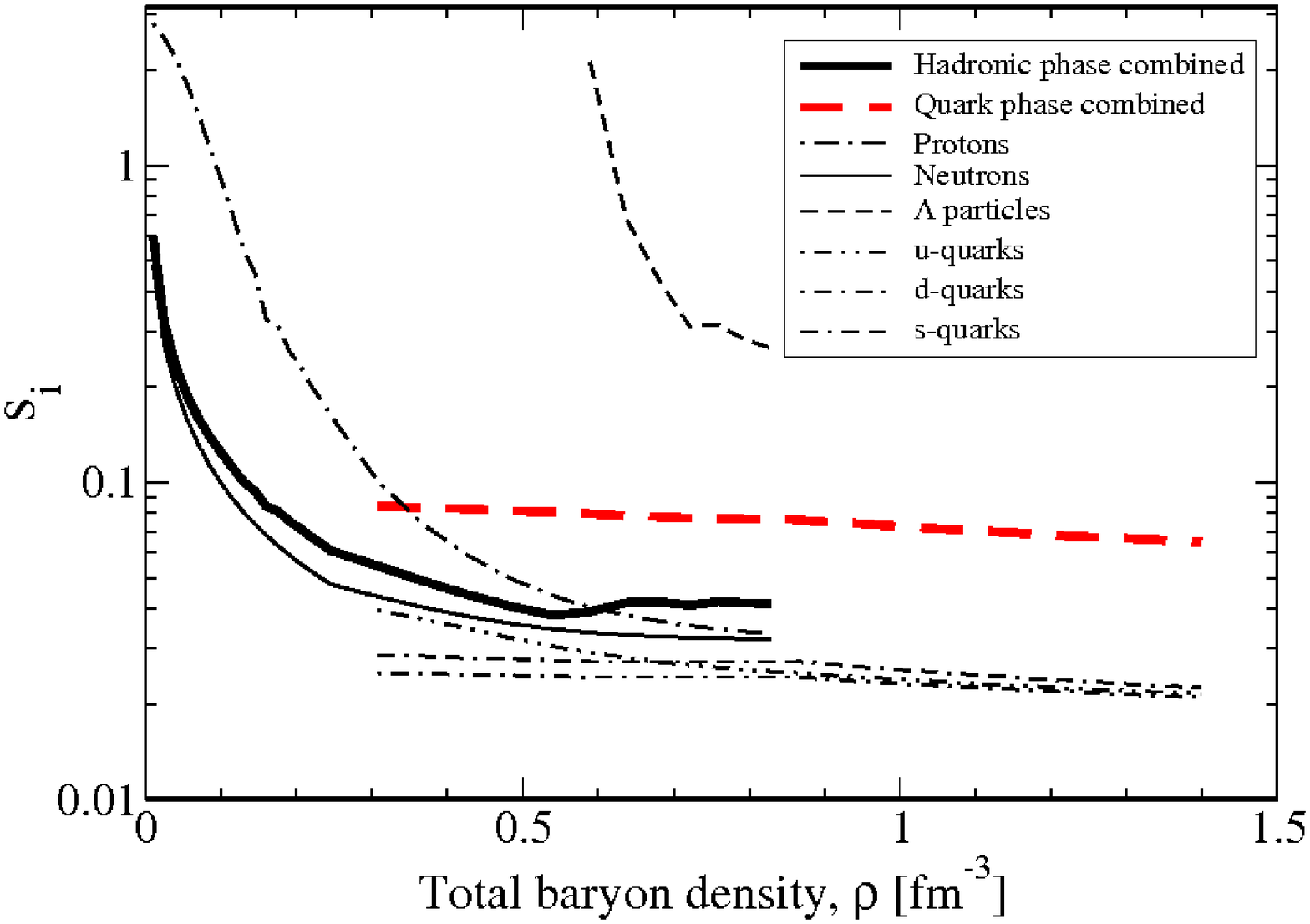}
\caption{Entropy per baryon at $T=1$~MeV in each phase, $s_\mathrm{H}$ and
  $s_\mathrm{Q}$, and per particle for each particle, $s_i$, as
  functions of total baryon density for the G$^{300}_{180}$ equation
  of state. Protons and $\Lambda$ particles have high $s_i$ because
  they are so rare, but contribute little to $s_\mathrm{H}$ for
  the same reason.}
\label{entropy}
\end{figure}

Figs.~\ref{composition} and \ref{entropy} show the resulting chemical
composition and entropy. In the quark phase we note
that u-quarks are suppressed initially giving the phase a negative net
charge, and that as expected the entropy per baryon -- though not per
quark -- is higher in the quark matter phase in the absence of any
pairing phenomena. We have checked numerically that in the absence of
pairing the entropy simply scales linearly with temperature within
the temperature range we shall need.

\begin{figure}[!htb]
\plotone{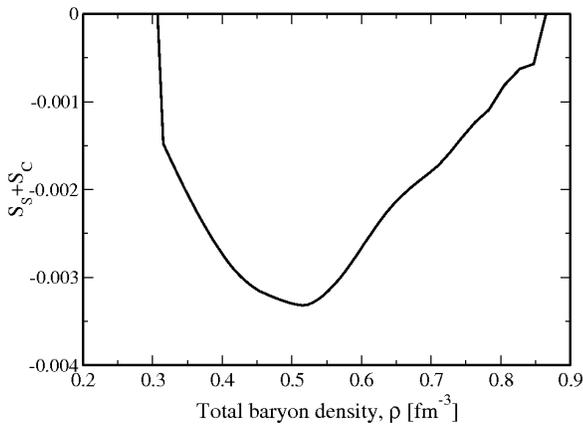}
\caption{Sum of surface and Coulomb entropy as a
      function of total baryon density for the G$^{300}_{180}$
      equation of state at $T=1$~MeV}
\label{ssurf}
\end{figure}

The sum of the surface and Coulomb energies has a maximum as a
function of density around $\rho=0.5-0.6 \textrm{ fm}^{-3}$ with a
corresponding minimum in the related entropy in Fig.~\ref{ssurf}. The
surface and Coulomb term in Eq.~(\ref{dsdt}) can therefore be either
positive or negative and either heat or cool the star
accordingly. The surface and
Coulomb contribution to the entropy is negative with the total entropy
remaining positive, which confirms that a structured phase has lower
entropy than an unstructured one.

We have calculated the structure of a sequence of stars with rotation
frequencies (as seen by an observer at infinity) between zero and the
limiting mass-shedding Kepler frequency for the G$^{300}_{180}$
equation of state. For this purpose we use the perturbative method of
\cite{Hartle:1967} and \cite{Hartle:1968} as implemented in the
numerical code developed by Weber which also solves self-consistently
for the general relativistic Kepler frequency $\Omega_\mathrm{K}$ --
see \cite{Weber:1999} for the derivation of $\Omega_\mathrm{K}$ and
\citep{Weber:1991,Weber:1992} for further details. The sequence has
constant total baryon number $A=1.87\,\times\,10^{57}$ and
nonrotating total gravitational mass $M=1.42$ M$_\odot$. The Kepler
frequency is then found to be $\Omega_\mathrm{K}=6168 \mbox{ rad
  s}^{-1}$ corresponding to a period of 1.02 ms.

\begin{figure}[!htb]
\plotone{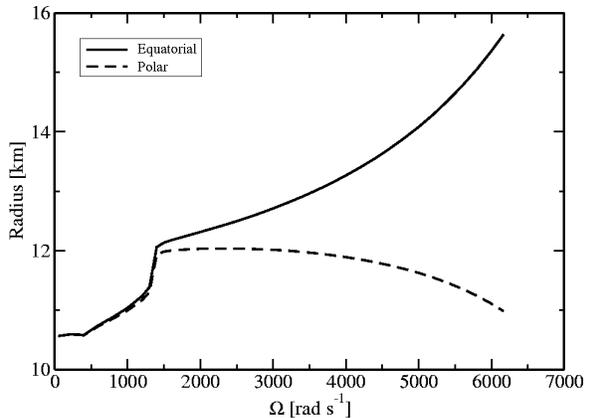}
\caption{Circumferential stellar radius in the equatorial and polar directions for
  rotating stars of total nonrotating gravitational mass $M=1.42$
  M$_\odot$.}
\label{OmegaR}
\end{figure}

\begin{figure}[!htb]
\plotone{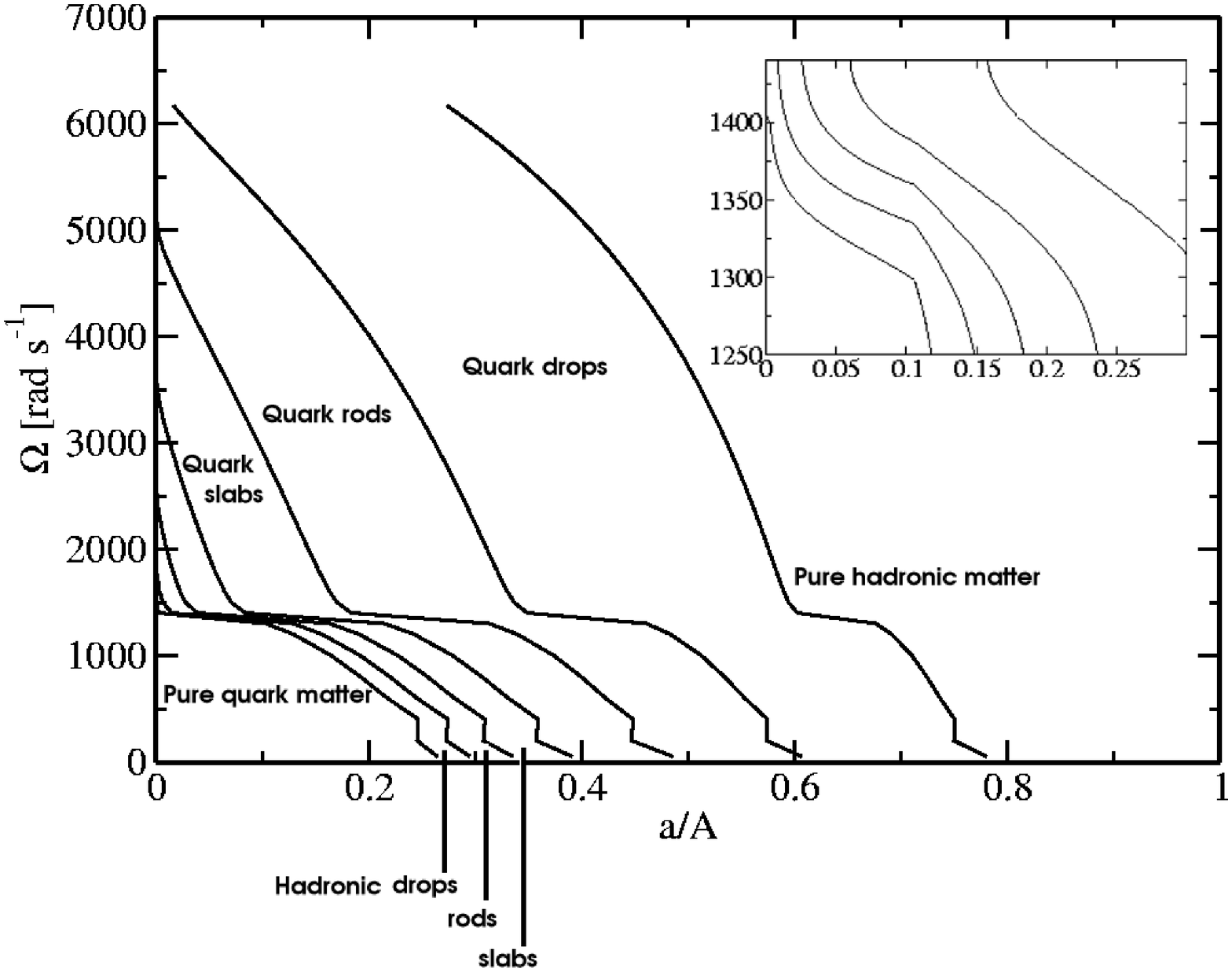}
\caption{Location of phase boundaries in stars of nonrotating mass
  $M=1.42$ M$_\odot$. The locations are given as the fractions of the
  total stellar baryon number contained within a surface of spatially constant
  density corresponding to each transition. The inset shows the region
  where the phase boundaries almost -- but not quite -- cross.}
\label{AObound}
\end{figure}

Figs.~\ref{OmegaR} and~\ref{AObound} show a few
properties of the models. The stars are significantly distorted by
rotation and increase their equatorial radius at the Kepler frequency
by half the nonrotating radius (Fig.~\ref{OmegaR}) while losing the
pure quark matter core at $\Omega=1400\mbox{ rad s}^{-1}$
(Fig.~\ref{AObound}). Since the stellar photon luminosity is
proportional to the surface area and must match the energy flux
emerging from the core, higher surface temperatures must also be
expected at low rotation frequencies for this reason alone. The
distortion from spherical symmetry depends on polar angle, however, and
above the frequency at which the quark matter core is lost the star
actually contracts in the polar direction. 

In Fig.~\ref{AObound} we show the location of the phase boundaries
between pure hadronic matter, the various geometries of the mixed
phase and the pure quark matter phase. Here we use as the free
variable the baryon number, $a$, contained within a surface on which
the density is uniform (i.e spatially but {\it not} temporally
constant). The Eulerian
density change $(\mathrm{d}\rho/\mathrm{d}\Omega)_r$ can be positive
or negative depending on location while the Lagrangian derivative
$(\mathrm{d}\rho/\mathrm{d}\Omega)_a$ is always negative, and so $a$
is a more convenient variable for some purposes. In Fig.~\ref{AObound} borders are shown at the densities which correspond to each
transition and $a$ is scaled to the total baryon number, $A$. We note that a large fraction of the star is
converted to quark matter as the star spins down. The pure quark
matter core appears below $\Omega=1400 \mbox{ rad s}^{-1}$ and
grows to eventually comprise $30$ \% of the stellar
gravitational mass and 26 \% of the stellar baryon number. 

\begin{figure}[!htb]
\plotone{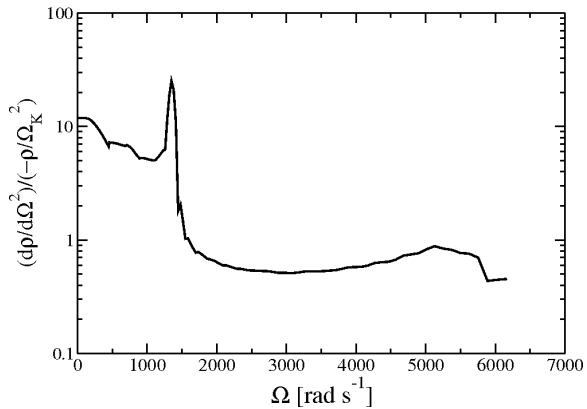}
\caption{Frequency derivative of density scaled to the estimate in
  Eq.~(\ref{drho}) at $a/A=0.33$ as a function of spin frequency. The
  estimate is best at high spin frequencies but holds within a factor
  of $~30$ in all stars.}
\label{OdrdO}
\end{figure}

Within the framework of \cite{Hartle:1967} the changes in pressure and
density with respect to a nonrotating star are second order effects
in the rotation frequency, and as discussed by \cite{Weber:1992} 
this remains true even at the limiting Kepler
frequency, $\Omega_K$, where the star would shed mass from the
equator. The Lagrangian density change during spindown can then be
reasonably approximated by the order of magnitude estimate (see also
\cite{Fernandez:2005})
\begin{equation}\label{drho}
\left(\frac{\partial \rho}{\partial \Omega^2}\right)_{a,A}\sim -\frac{\rho}{\Omega_\mathrm{K}^2}
\end{equation}
Fig.~\ref{OdrdO} shows how the numerically calculated
spin down compression rate compares to this estimate in the mixed
phase -- we have checked it at other positions as well. For stars with
high spin frequencies and no pure quark matter core Eq.~(\ref{drho}) is
a reasonable approximation although an overestimate in some
regions. It is off by approximately a factor
of 0.1 below the frequency where the pure quark matter core
forms.

Eq.~(\ref{drho}) can be used to give a rough estimate of
the effects of the additional entropy production related to density
change in Eq.~(\ref{dsdt})
\begin{eqnarray}\label{dsdt2}
T\frac{\mathrm{d}s}{\mathrm{d}t}&=&\frac{c_\mathrm{v}}{\rho}\frac{\mathrm{d}T}{\mathrm{d}t}+T\frac{\mathrm{d}s}{\mathrm{d}\rho}\frac{\mathrm{d}\rho}{\mathrm{d}t}\\
&&\simeq\frac{c_\mathrm{v}}{\rho}\frac{\mathrm{d}T}{\mathrm{d}t}-T\frac{\mathrm{d}s}{\mathrm{d}\rho}\frac{\rho}{\Omega_\mathrm{K}^2}\times2\Omega{\dot\Omega}\;.\nonumber
\end{eqnarray}
Taking the entropy per particle from Eq.~(\ref{s}) this implies that the
bulk terms in Eq.~(\ref{dsdt}) are of the order of
\begin{align}
T\frac{\mathrm{d}\rho}{\mathrm{d}t}\frac{\mathrm{d}s}{\mathrm{d}\rho}&\simeq\frac{2}{3}\frac{\Omega{\dot\Omega}}{\Omega_\mathrm{K}^2}Ts\nonumber\\
&\simeq-10^{-15}T_9^2\left(\frac{B}{10^{14}\mbox{
    G}}\right)^2\\ 
&\times\left(\frac{\Omega/6000}{\mbox{ rad
    s}^{-1}}\right)^4\left(\frac{\rho}{\mbox{fm}^{-3}}\right)^{-\frac{1}{3}}\mbox{erg s}^{-1}\;,\nonumber
\end{align}
where we have taken $\Omega_\mathrm{K}=6000\mbox{
  rad s}^{-1}$, assumed a standard dipole model for the spindown so
  $\dot{P}=(B_{19}/3.2)^2 P^{-1}$ with the spin period, $P$, measured
  in seconds and the magnetic field, $B$ in units of $10^{19}\mbox{
  G}$. As a rough estimate the total heating power from the bulk terms
  in Eq.~(\ref{dsdt}) in the absence of
  pairing and for a star of constant density and temperature with a baryon
  number of $10^{57}$ is then of the order of
\begin{eqnarray}
W_\mathrm{Bulk}&\sim &10^{42}T_9^2\left(\frac{B}{10^{14}\mbox{
    G}}\right)^2\left(\frac{\Omega/6000}{\mbox{ rad
    s}^{-1}}\right)^4\nonumber\\
&&\times\left(\frac{\bar{\rho}}{\mbox{ fm}^{-3}}\right)^{-\frac{1}{3}}\mbox{erg s}^{-1}\;. \label{wbulk}
\end{eqnarray}

This includes only the bulk terms of Eq.~(\ref{dsdt}), but on general
grounds one would expect this to be a reasonable approximation of the
total additional entropy production. The latent heat has the same
basic origin -- the release of entropy with changing density -- and is
therefore expected to be of the same order of magnitude as the bulk
terms in Eq.~(\ref{wbulk}). In the following section we shall find that
this is justified under most, but not all, circumstances.

The estimated heating in Eq.~(\ref{wbulk}) is to be compared with the
neutrino and photon luminosities which in the absence of pairing
phenomena can roughly be estimated as \citep{Page:2006a}
\begin{eqnarray}
L_\nu^\mathrm{slow}\sim 10^{40}T_9^8\mbox{ erg s}^{-1}\\
L_\nu^\mathrm{fast}\sim 10^{45}T_9^6\mbox{ erg s}^{-1}\\
L_\gamma\sim 10^{33}T_9^2\mbox{ erg s}^{-1}
\end{eqnarray} 
where 'slow' refers to stars dominated by relatively inefficient
neutrino emission processes such as the modified Urca cycle,
bremsstrahlung or the pair breaking and formation process while 'fast'
refers to stars dominated by the highly efficient direct Urca
process. 

From Eq.~(\ref{wbulk}) we then see that while it may be possible to find
combinations of temperature, magnetic field and rotation frequency for
which the additional entropy production dominates, such configurations
may be difficult to realize in nature. For instance, in order for the
heating term in Eq.~(\ref{wbulk}) to dominate the neutrino luminosities
the star must be a hot millisecond magnetar. Such an object would be
very short lived if it could be created in nature at all. For the
heating term to dominate the photon luminosity -- which is the more
relevant term at late times and low temperatures -- extremely high
magnetic fields would again be required assuming a dipole model for
the spin down. Such conditions -- if they were to be realized -- would
be very short lived and therefore the more difficult to observe.

From these simple estimates we then expect that the release of entropy
related to changing bulk density with spin down, the latent heat of
phase transitions and changes in the surface and Coulomb energy of
mixed phases (smaller still) will have little impact on the thermal
evolution of compact stars and be difficult to observe.

\section{Spherical Isothermal Cooling} \label{calculations}
Estimates cannot replace detailed calculations, and the ones discussed
above further ignore all effects of superfluidity, the changing
chemical composition and the rapid changes in structure at certain
frequencies. We therefore proceed in this section to explore these
effects through numerical calculations taking the simplest possible
approach to solve the thermal balance in Eq.~(\ref{balance2}).

\subsection{Numerical Setup}

Neutron stars are commonly taken to be isothermal
after an initial thermal relaxation lasting 10-100 years in the
sense that the redshifted temperature $T^\infty=Te^\Phi$ is constant
below a thermally insulating layer in the outer crust \citep{Gudmundsson:1983}. Since we intend only to investigate the late thermal
evolution of neutron stars, and as we do not expect the heating terms
discussed here to change the isothermality by disturbing the thermal
balance of any part of the star with its surroundings, we shall work
within this same approximation. To show the full range of effects,
plots in the following also include young stars not expected to be
isothermal and our results should only be taken as indicative at such
early times.

Assuming the star to be thermally relaxed, so the redshifted
temperature $T^\infty=Te^\Phi$ is constant below the outer crust, we then
integrate Eq.~(\ref{balance2}) to get a simplified equation for global
thermal balance
\begin{equation}\label{sphbal}
\frac{dT^\infty}{dt}=\frac{1}{C}\left(W^\infty-L_\nu^\infty-L_\gamma^\infty\right)
\end{equation}
with
\begin{eqnarray}
C&=&\int \mathrm{d}a\, T\frac{\partial s}{\partial T} \label{C}\\
W^\infty&=&-\int \mathrm{d}a\, e^{\Phi}T\frac{\partial
  s}{\partial\rho}\frac{\partial\rho}{\partial\Omega}{\dot \Omega}   \label{Wdef}
\\ 
L^\infty_\nu&=& \int \mathrm{d}a\, \frac{\epsilon_\nu}{\rho} e^{2\Phi}\\ 
L^\infty_\gamma&=&4\pi R^2\sigma (T_\mathrm{S}^{(0)})^4F(B)e^{2\Phi_s}\label{Lgamma}\\
 T_{\mathrm{S}_6}^{(0)}&=&
 g_{14}^{1/4}[(7\zeta)^{2.25}+(\zeta/3)^{1.25}]^{1/4}\\\label{Ts}
\zeta&=&T_9-0.001\;g_{14}^{1/4}\sqrt{7T_9}
\end{eqnarray}
where $\epsilon_\nu/\rho=Q_\nu$ is the neutrino emissivity, the
surface temperature is given in units of $10^6$~K, 
$T_\mathrm{S_6}=T_\mathrm{S}/10^6$~K and $T_\mathrm{S}^{(0)}$ is
surface temperature at zero magnetic field strength. We have
used a fit appropriate for iron envelopes to the relation between the
surface temperature, $T_\mathrm{S}$, and the inner temperature below
the heat-blanketing envelope, $T$, in which $g_\mathrm{{14}}$ is the
surface gravity in units of $10^{14}$ cm s$^{-2}$
\citep{Potekhin:1997,Potekhin:2001}. 

The heat conductivity
in the crust, which derives mainly from the motion of electrons, becomes
anisotropic in a strong magnetic field \citep{Potekhin:2001, Geppert:2004}. It is slightly enhanced in the
direction along the magnetic field lines and strongly suppressed in
the direction orthogonal to the field lines. The heat blanketing
relation and the surface temperature
then becomes nonuniform,
$T_\mathrm{local}(B,\theta,g,T)=T_\mathrm{S}^{(0)}(g,T)\chi(B,\theta,T)$.
The photon luminosity must therefore be found by integrating the locally
emerging flux and it will depend on magnetic field strength,
$L_\gamma(B)=L_\gamma(0)F(B)$. For this purpose we use the fits given
by \cite{Potekhin:2001} for $\chi(B,\theta,T)$ and $F(B)$ in a dipole
magnetic field. These are good for magnetic fields below $10^{16}$~G,
interior temperature between $10^7$~K and $10^9$~K and surface
temperature above $10^5$~K. $F(B)$ reaches a minimum of $\sim 0.7$
around $B=10^{13}$~G and grows to about 2 at $B=10^{15}$~G for a star
of internal temperature $T=10^9$~K. The location of the minimum moves to
lower magnetic field strength for lower internal temperature and the
change in photon luminosity can delay or accelerate the cooling at
late times accordingly. The figures in Sect.~\ref{numericalresults}
show the effective surface temperature at infinity
$T_\mathrm{S}^\infty=T_\mathrm{S}^{(0)}F(B)^{1/4}e^\Phi$. A strong
magnetic field in the inner crust can also lead to anisotropic heat
flow there (see e.g. \cite{Geppert:2004} and \cite{Aguilera:2007}),
but these effects cannot be included here.\\

For given assumptions about the neutrino emissivity and superfluid
gaps Eq.~(\ref{sphbal}) is solved along with the spin down model to give
the surface temperature at infinity as a function of time, rotation
frequency and magnetic field strength. At each time step Eqs.~(\ref{C}) to~(\ref{Ts}) are solved to update the stellar structure at the
appropriate rotation frequency.\\

Our equations of thermal balance assume spherical symmetry. This
effectively treats rotation as a perturbation whose only effects are
to change the size and chemical composition of the stars and to
provide additional terms in the thermal balance. This approach
ignores the effects of nonradial heat flows and nonspherical
perturbations of the metric which can be significant when the star is
not spherically symmetric at very high spin frequencies (see
Fig.~\ref{OmegaR}). Two-dimensional cooling calculations are beyond
our scope, but they have been performed at constant rotation frequency
by \cite{Schaab:1998a} and \cite{Miralles:1993}. For stars rotating at
a large fraction of their Kepler frequency these authors found
significant effects on the thermal evolution during the nonisothermal
epoch, and polar temperatures up to 31\% higher than the equatorial
temperatures even for internally isothermal stars. The recent work of
\cite{Geppert:2004, Aguilera:2007, Aguilera:2008} further includes the
influence of large magnetic fields in the crust and Joule heating by
magnetic field decay in two dimensions. This results in nonradial
heat flow and significant heating from magnetic field decay. We focus
on the effects of deconfinement during spin down however and shall not
pursue these aspects here.

Unless otherwise stated we shall use the equatorial radius in
Eqs.~(\ref{Lgamma}) to~(\ref{Ts}). We have performed calculations with
the polar radius too and will show one such example. At high rotation
frequency the surface temperature is then higher and responds less to
changes in rotation frequency. We expect results in a two-dimensional
code would be intermediate between these and such a calculation would
be a significant improvement on what is presented here. With this in
mind we use the equatorial radius to show the most pronounced effects
of spin down. However, we also note that the strongest effect we have
found from spin down stems from the formation of a pure quark matter
core. At the rotation frequency where this happens the polar and equatorial
radii are very similar (see Fig.~\ref{OmegaR})
and a spherical approximation is reasonable.\\

Since the stellar moment of inertia $I$ and radius also change with rotation
frequency we modify the spin down model described in the previous
section to include such effects. The spin down is then determined by
\citep{Glendenning:2000}
\begin{align}
\dot{\Omega}&=-\frac{K}{I}\left[1+\frac{I'\Omega}{2I}
  \right]^{-1}\Omega^m, \quad
  I'=\frac{\mathrm{d}I}{\mathrm{d}\Omega}\label{Odot}\\
K&=(2/3)R^6 B^2\sin^2{\alpha}
\end{align}
where $\alpha$ is the inclination angle between the magnetic axis and
the rotation axis and $m-1$ is the multipolarity -- usually $m=3$ for
magnetic dipole braking or $m=5$ for gravitational quadrupole
radiation. We take $\sin^2\alpha=1$ and use the canonical value $m=3$ for a dipole spin down.\\

As for the neutrino emissivities and superfluid properties of the
neutron star we are interested in the most important effects only and
aim for transparency in the model. In the quark phase we
include the direct and modified Urca cycles as well as bremsstrahlung
using the emissivities in \cite{Blaschke:2000}. The $us$-branch of the
direct Urca cycle fulfills the triangle inequality,
\begin{equation}
k_{\mathrm{F}_s}<k_{\mathrm{F}_u}+k_{\mathrm{F}_e} 
\end{equation}
where $k_{\mathrm{F}_i}$ is the Fermi momentum of the strange-quark, up-quark and electron respectively. It is thus allowed throughout the
quark phase, but its $ud$-counterpart is forbidden as it cannot
conserve both energy and momentum. This difference stems from the
assumed strange quark mass of 150 MeV. 

In the hadronic phase we again include the direct and modified Urca
cycles as well as bremsstrahlung using the emissivities given in
\cite{Yakovlev:2001}. We ignore the effects of hyperons here because
of their low abundance. In neutron stars the direct Urca reaction is
often allowed only at very high densities because it cannot fulfill
both energy and momentum conservation unless the proton
fraction is above the value where both charge neutrality and the
triangle inequality can be observed \citep{Lattimer:1991}. It is
critical to note that this is not so in hybrid stars with a mixed
phase. The mixed phase is possible precisely because charge neutrality
does not have to be observed locally, and it is favorable because the
hadronic phase lowers its nuclear symmetry energy by increasing the
proton fraction (Fig.~\ref{composition}). Hence the hadronic direct
Urca cycle is active in the mixed phase, and as it is about three orders
of magnitude faster than the $us$-branch of the quark direct Urca
cycle it controls the cooling of hybrid stars with a mixed phase
unless reduced by pairing effects.

Additional neutrino processes should be included in a more
sophisticated treatment -- particularly in the crust -- but as stated
above we are mainly interested in the general properties of the
additional entropy production and the changing chemical composition
and we shall here leave out such terms.\\

If there is any attractive interaction among particles in a degenerate
Fermi system they will pair, and the resulting superfluidity has
important consequences for the thermal properties of neutron star
matter and the thermal evolution of neutron stars. Pairing in general
delays the cooling because it suppresses most of the neutrino
emissivities and enhances the heat capacity at temperatures just below
the critical. However it also opens additional neutrino emission
processes, suppresses the heat capacity far below the critical
temperature and enhances the thermal conductivity, and it may
therefore also accelerate the cooling at certain epochs (see
e.g. \cite{Yakovlev:1999} for a detailed account).

There is considerable uncertainty concerning the relevant superfluid
phase in quark matter -- see \cite{Alford:2007a} for a comprehensive
review of the effects of pairing among quarks and the possible range
of phases. For this first investigation of consequences for the
thermal evolution of a quark-hadron phase transition during spin down
we assume a simplified but physically transparent model for quark
matter pairing which essentially corresponds to pairing in the
color-flavor locked phase \citep{Alford:1999,Alford:2001}. In this
model each flavor participates equally, the gap $\Delta_{Q,0}$ is
independent of density and the critical temperature is
$T_c=0.72\Delta_{Q,0}$ (see \cite{Schmitt:2002}). We shall consider a
wide range in $\Delta_{Q,0}$ partly because it is interesting for the
present calculation, partly because very low gaps may be realized for
high strange quark mass as this implies unequal quark Fermi momenta in
the unpaired phase. The color-flavor locked phase is generally
understood to be so strongly suppressed in all its thermal properties
as to be virtually inert with respect to cooling. Since we shall also
consider small gaps and since the latent heat derived from the phase
transition depends strongly on pairing in the quark phase this will not
always apply here, but for strong quark pairing the thermal evolution
is generally controlled by the hadronic phase. We further note that as
shown by \cite{Jaikumar:2002} (and stressed by \cite{Alford:2007a})
Goldstone modes and their related neutrino emissivities and heat
capacities are not exponentially suppressed, scaling instead as
$T^{15}$ and $T^3$ respectively. They will therefore dominate in
color-flavor locked quark matter at low temperatures, but they are
exceedingly small and do not influence our results.\\

The effects of superfluidity relevant for our purpose are that below
the critical temperatures it suppresses the neutrino emissivities and
entropies of the participating particles while also allowing
additional neutrino emission through Cooper pair breaking and
formation. The specific heat is enhanced just below the critical
temperature and suppressed at lower temperatures. These effects are
incorporated through control functions, $R(T,T_c)$, which multiply the
relevant quantities. They depend only on the pairing channel,
temperature and gap size, and may therefore be calculated given
expressions for the gap size momentum dependence.\\

In the quark matter phase we use the simplest possible control
functions and suppress the direct Urca emissivity by a factor of
$e^{-\Delta_\mathrm{Q}/T}$ and the modified Urca and bremsstrahlung
emissivity by a factor of $e^{-2\Delta_\mathrm{Q}/T}$, where
(following \cite{Steiner:2002})
$\Delta_\mathrm{Q}=\Delta_{Q,0}\sqrt{1-(T/T_c)^2}$ is the pairing gap
at the local temperature, $T=T^\infty e^{-\Phi}$.

The specific heat and entropy of superfluid particles are also
modified at temperatures below the critical temperature. In the
quark matter phase we use the fit to the $^1S_0$ control function for
the specific heat given by \cite{Yakovlev:1999}. We further fit the results of
\cite{Muhlschlegel:1959} for the entropy suppression to obtain
\begin{align}\label{ssf}
s_\mathrm{sf}&=0.95\,s_0\times(T/T_\mathrm{c})e^{-T_\mathrm{c}/T}\nonumber\\
&\times\left[0.43+3.82(T/T_\mathrm{c})-1.41\left(T/T_\mathrm{c}\right)^2\right]\,.
\end{align} 
Where $s_0$ is the nonsuperfluid entropy. This expression is applied
to reduce the entropy of both quarks and hadrons (thus here neglecting
the difference between different pairing states of neutrons). The
entropy difference between the superfluid and the normal phase depends
on temperature as $\Delta s\propto - (1-T/T_\mathrm{c})$ close to the
critical temperature \citep{Landau:1980} with $s_0$ becoming
negligible below 0.2$T_\mathrm{c}$. This has the effect of gradually
turning the latent heat in Eq.~(\ref{dsdt}) from a cooling term into a
heating term, but it also suppresses the bulk terms in
Eq.~(\ref{dsdt}).

In the absence of detailed calculations of control functions for the
entropy of superfluid quark matter phases in the literature we have
employed Eq.~(\ref{ssf}) in the quark matter phase though it is
strictly relevant only for BCS superfluidity. Eq.~(\ref{ssf}) is,
however, consistent with the exponential suppression of entropy below
the critical temperature which would be expected in any such
calculation, and we do not expect that our results are sensitive to
this choice.

Pairing further opens the important possibility of neutrino emission
by Cooper pair breaking and formation in the quark phase and we
include the associated emissivity as described in \cite{Jaikumar:2001}
using the fit to the $^1S_0$ control function in
\cite{Yakovlev:1999}.\\

Among the nucleons pairing is predicted for neutrons in the $^1$S$_0$
and $^3$P$_2$ channels and for protons in the $^1$S$_0$ channel. The
gaps can be calculated self-consistently but results are uncertain and
vary greatly both in terms of the maximum gap size and in terms of
density dependence -- see e.g. \cite{Lombardo:2001,
  Page:2004,Page:2006a} for collections of these results. We shall
here use gaps with a phenomenological momentum dependence as suggested
by \cite{Kaminker:2001}, \cite{Andersson:2005} and recently
\cite{Aguilera:2007}
\begin{align}
\Delta(k_{F,N})=\Delta_0&\frac{(k_{F,N}-k_0)^2}{(k_{F,N}-k_0)^2+k_1^2}\\
&\times\frac{(k_{F,N}-k_2)^2}{(k_{F,N}-k_2)^2+k_3^2}\label{deltaofk}
\end{align}
where $k_{F,N}$ is the Fermi momentum of the relevant nucleon,
$N=n,p$. Our choices for the parameters $\Delta_0$ and $k_{i}$ correspond to sets $a$, $e$ and $h$ of
\cite{Aguilera:2007} (see that work for references to the model
calculation underlying these fits). Eq.~(\ref{deltaofk}) is valid only
in the range $k_0<k<k_2$ with $\Delta(k_{F,N})=0$ outside this range,
and where the gaps for $^1$S$_0$ and $^3$P$_2$ pairing of neutrons are
both nonzero we use the largest of the two gaps. We further use the
relations between critical temperature and pairing gap listed by
\cite{Yakovlev:1999}.\\

In the hadronic phase neutrons and protons may pair simultaneously in
different channels and so the resulting calculations for the control functions can be quite involved. We use the fits to numerically
calculated control functions for each process and each pairing channel
compiled in \cite{Yakovlev:1999} (or
\cite{Yakovlev:2001} -- see these works for detailed
references). Where protons and neutrons pair simultaneously we use the
combined factors listed in these works if available and if not then
the smallest of the two independent control functions. 

The hadronic pair breaking and formation process can be very powerful
and may dominate the thermal evolution when pairing first sets in, but
it was recently shown by \cite{Leinson:2006} to be strongly suppressed
in the singlet state channel by approximately a factor of $10^{-6}$
relative to the $^3$P$_2$ channel (see also the recent work of \cite{Steiner:2008}). In the hadronic phase we therefore
include pair breaking and formation only for neutrons at high
densities where they pair in the $^3$P$_2$ channel and for this we
again use the emissivity and control function given in
\cite{Yakovlev:1999,Yakovlev:2001}.
 
\subsection{Numerical Results}\label{numericalresults}
\begin{figure}[!htb]
\plotone{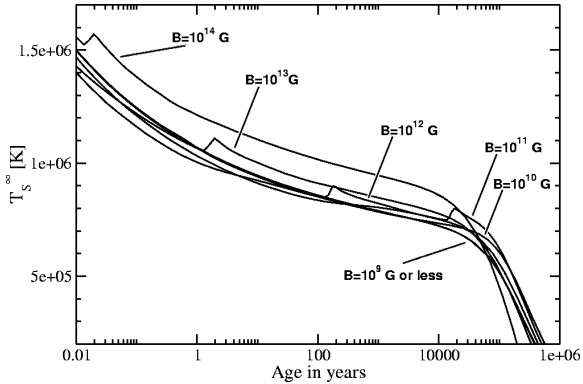}
\caption{The effect of spin down compression on cooling curves for
  stars with initial spin frequency $\Omega_\mathrm{start}=6000 \mbox{ rad
  s}^{-1}$ and constant magnetic field as labeled for each curve. The
  quark core is assumed superconducting with $\Delta_{Q,0}=10$
  MeV. The increasing quark matter fraction produces slower cooling
  and a jump in temperature with the appearance of the pure quark
  matter phase.}
\label{Tt}
\end{figure}

\begin{figure}[!htb]
\plotone{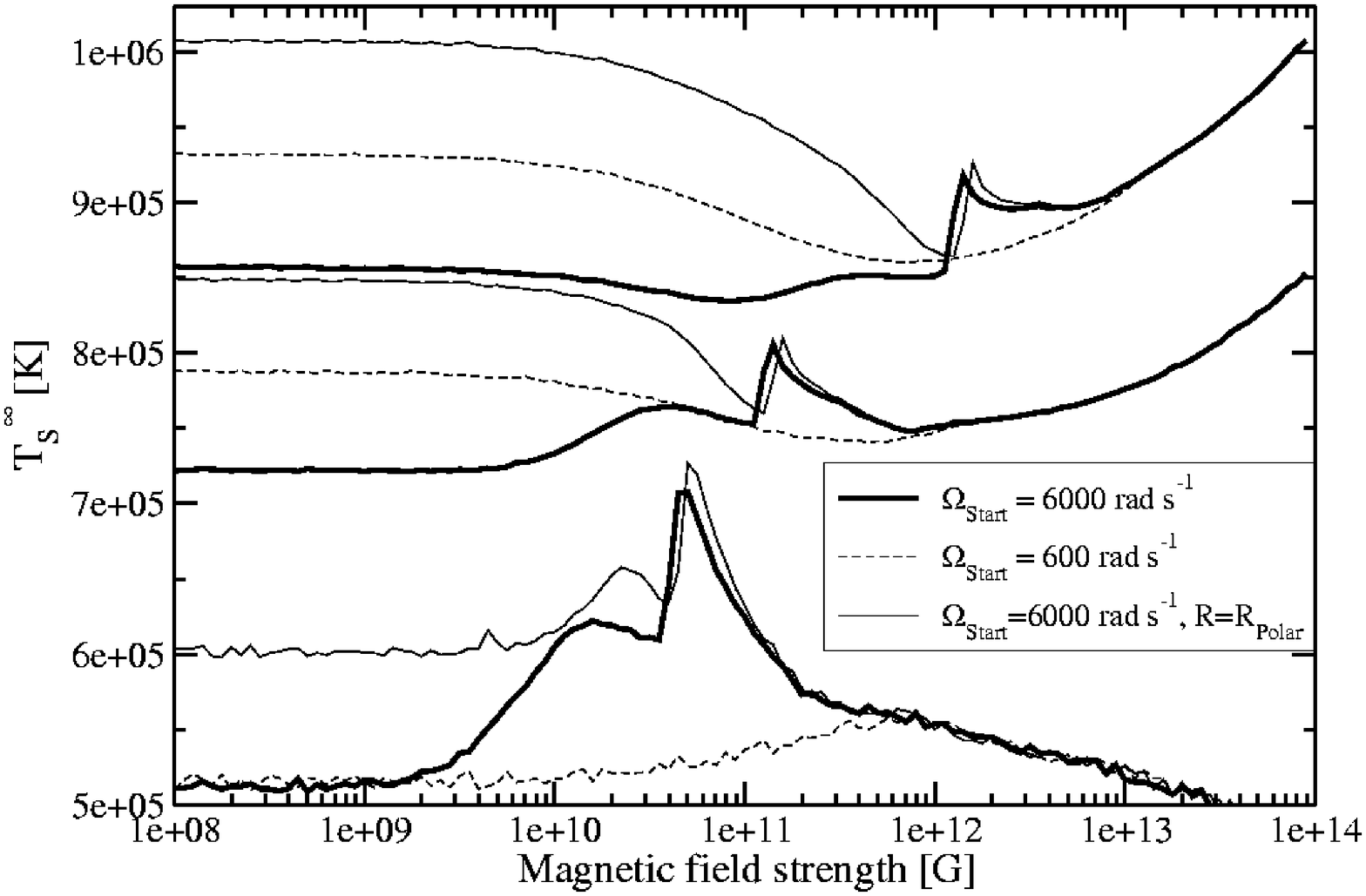}
\caption{Variation of surface temperature at infinity with magnetic
  field for stars at ages $10^2$, $10^4$ and $10^5$ years from
  above. Thick lines have initial spin frequency $\Omega_\mathrm{start}=6000
  \mbox{ rad s}^{-1}$ while thin dashed lines start with
  $\Omega_\mathrm{start}=600 \mbox{ rad s}^{-1}$. Thin continuous lines use the
  polar rather than the equatorial radius. The quark core is assumed
  superconducting with $\Delta_{Q,0}=10$~MeV. We note a slow increase
  in temperature for old stars and a sudden jump at the magnetic field corresponding
  to significant spin down at specific ages. The suppression in
  temperature at low magnetic field and increase at high field is due
  to the effects of the magnetic field of the heat blanketing relation.}
\label{BT}
\end{figure}

\begin{figure}[!htb]
\plotone{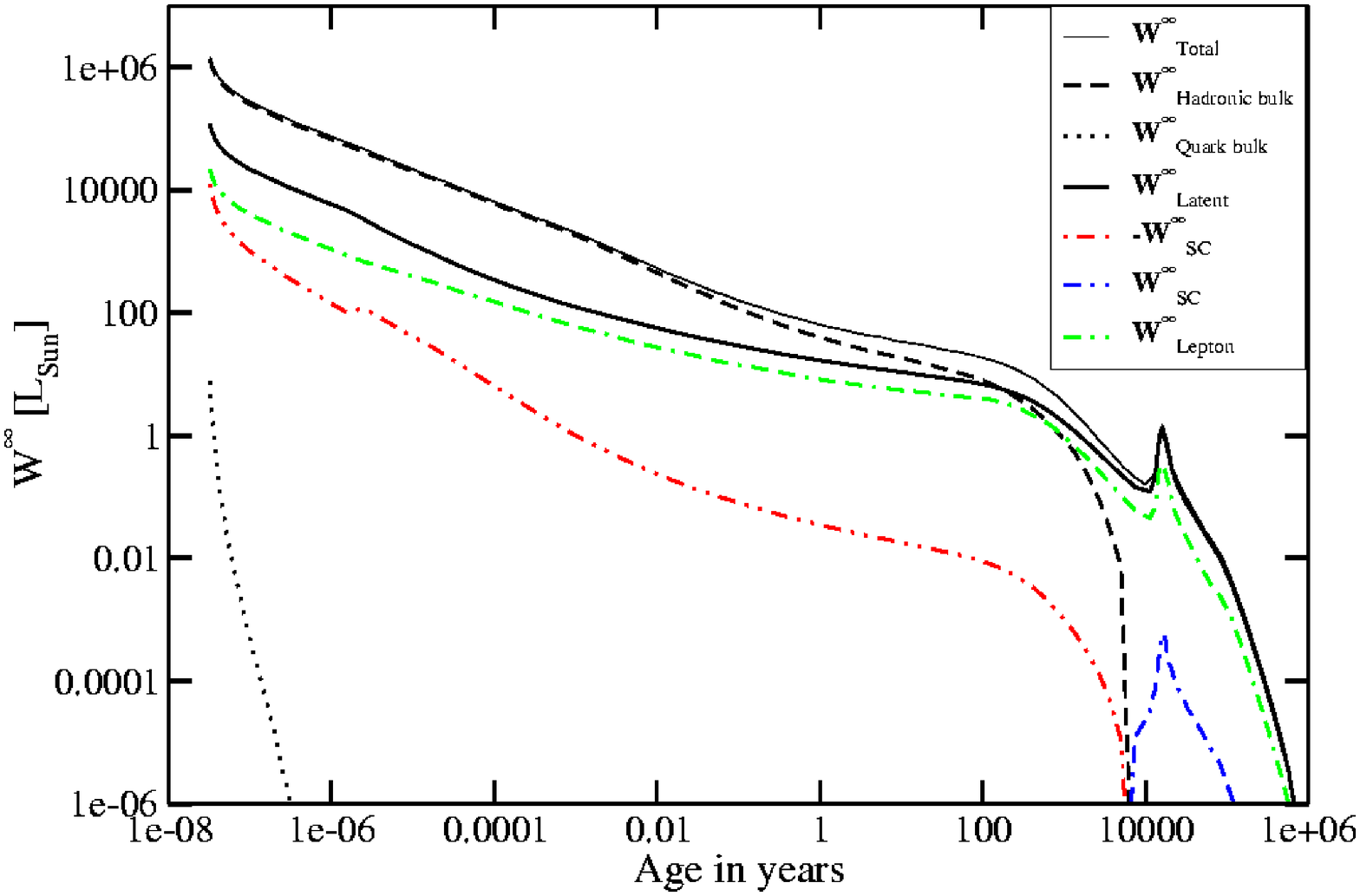}
\caption{Numerical value of the entropy production as defined in
  Eq.~(\ref{Wdef}) as a function of time for a star with magnetic field
  $B=10^{11}$~G and quark pairing gap $\Delta_{Q,0}=10$~MeV. $W_\mathrm{Tot}$ is split into
  contributions with different physical origins as explained in the
  text. The surface and Coulomb term changes sign and the negative of
  this term is shown as well. The latent heat and bulk terms are dominant, and the spike is
  caused by the appearance of a pure quark matter phase.}
\label{Wt}
\end{figure}

\begin{figure}[!htb]
\plotone{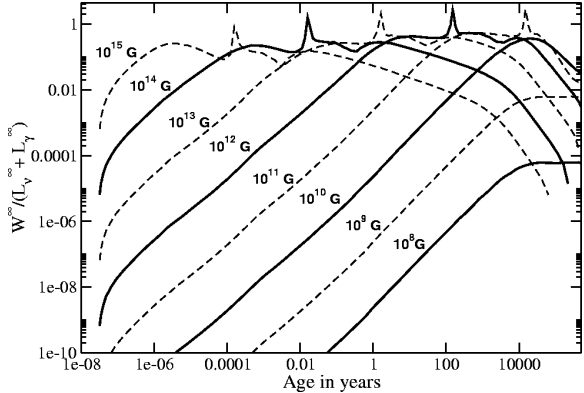}
\caption{The total entropy production scaled to the combined neutrino
  and photon luminosity for stars with quark paring gap
  $\Delta_{Q,0}=10$~MeV and magnetic fields as indicated just above
  each curve. The different line styles are meant only to help guide
  the eye. Note that at the peaks reaching above 1 the entropy
  production actually dominates the cooling terms and the
  temperature increases. Also note that this is only possible for
  stars changing structure rapidly with the introduction of a pure
  quark matter core.}
\label{Wtscaled}
\end{figure}

\begin{figure}[!htb]
\plotone{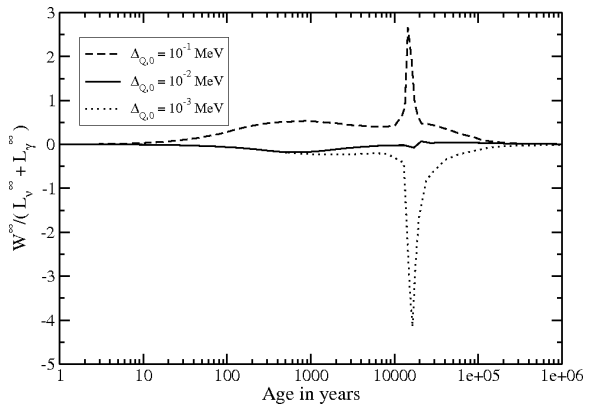}
\caption{The total entropy production scaled to the combined neutrino
  and photon luminosity for stars with quark paring gap as indicated
  in the legend and $\Omega_\mathrm{start}=6000 \mbox{ rad s}^{-1}$. The
  magnetic field is constant at $10^{11}$~G. The entropy production
  changes sign and becomes negative for low pairing gaps and thus acts
  as a powerful cooling term during some epochs.}
\label{WtDelta}
\end{figure}

\begin{figure}[!htb]
\plotone{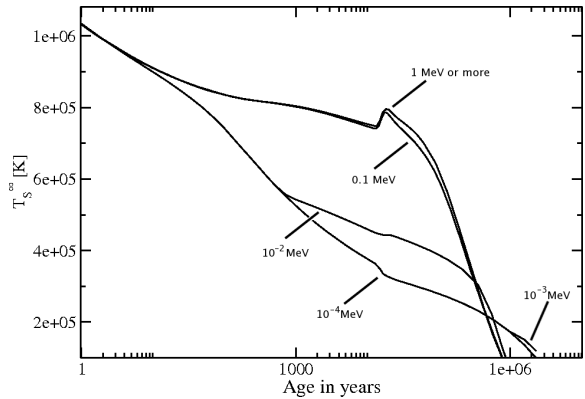}
\caption{The effect of spin down compression on cooling curves for
  stars with initial spin frequency $\Omega_\mathrm{start}=6000 \mbox{ rad
  s}^{-1}$ and quark pairing gap $\Delta_{Q,0}$ as labelled for each
  curve. The magnetic field is constant at $10^{11}$~G}
\label{TtDelta}
\end{figure}

Figures~\ref{Tt} to~\ref{TtDelta} explore consequences for the
thermal history of hybrid stars of including time dependent structure
and entropy production by spin down
compression in the cooling calculations with the approximations discussed above. They show the
surface temperature and additional entropy production at infinity as
functions of time, magnetic field and quark pairing gap energy.

In Fig.~\ref{Tt} we show the thermal evolution of stars with quark
pairing gap $\Delta_{Q,0} = 10$~MeV, a range of magnetic fields and
initial spin period around 1~ms. The extremely rapid initial rotation
is interesting because it means the stars start out with no pure quark
matter core and undergo drastic changes in structure as a pure quark
matter core develops around $\Omega=1400$ rad s$^{-1}$. We see this in
Fig.~\ref{Tt} as a short period of increasing temperature at the time
corresponding to this angular velocity for a given magnetic
field. During this short period the heating term from the additional
entropy production actually dominates which is shown in more detail in
Fig.~\ref{Wtscaled} discussed below. Further, the changing neutrino
emissivity and heat capacity with the introduction of more quarks in
the star and the reduction in stellar surface area as the pure quark matter
core develops all affect the thermal evolution. This implies somewhat
slower cooling -- with our specific choice of parameters -- and higher
surface temperature. For the internal temperature, $T$, we have found
essentially a transition between two otherwise similar cooling tracks;
one for stars with no pure quark matter core and one for stars with a
fully developed quark matter core. The effective surface temperature
also depends on magnetic field strength through $F(B)$ however. For
this reason such an easy interpretation is difficult from
Fig.~\ref{Tt} alone.\\

The magnetic field strength determines the spin down and affects the
heat blanketing relation. Fig.~\ref{BT} shows how the temperature varies at
specific ages with the magnetic field strength (which is kept constant in time
itself). This is shown for initial spin frequencies of 6000 rad
s$^{-1}$ and 600 rad s$^{-1}$ and using the polar radius in addition to
the equatorial when calculating the surface temperature.

For low initial spin frequency
we find no significant effects of spin down in Fig.~\ref{BT}. The
change in temperature at a specific age
with increasing magnetic field is due to the effects of the magnetic
field on the heat blanketing relation and photon luminosity. If these
effects were left out the curves for low initial spin frequency would
be almost constant. 

If the initial rotation frequency is sufficiently high to exclude the
pure quark matter phase it is a different matter. A strong magnetic
field initially suppresses the surface temperature of young stars with
high rotation frequencies, but it then gives a sharp jump at the
magnetic field strength which ensures that a pure quark matter core is
formed at a particular age. For older stars there is also a slight
increase in surface temperature before the formation of a pure quark
matter core when the equatorial radius is used. If the effects of the
magnetic field on the heat blanketing relation were left out the
curves at high initial spin frequency become nearly constant for
$B>10^{13}$~G and for fields too weak for spin down to set in at the
particular age. In this case there would still be a jump in
temperature at the magnetic field strength corresponding to formation
of a pure quark matter core for a specific age however.

Replacing the equatorial radius with the polar in Eqs.~(\ref{Lgamma})
and~(\ref{Ts}) (thin continuous lines in Fig.~\ref{BT}) we find higher
surface temperatures before the introduction of pure quark matter in
the core and still a clear jump in temperature after. The gradual
increase in temperature before the jump found in the other curves is
absent except for old stars. This can be understood if we remember
that the polar radius is smaller than the equatorial and less
sensitive to spin down at high rotation frequencies. We have checked that a similar pattern may
be seen in plots of temperature versus time using the polar
radius. We expect that full two-dimensional calculations would
give results which are at high rotation frequencies intermediate
between what we have found using a spherical approximation. The similarity between curves with polar and equatorial radius
at the jump in temperature shows that our calculations are insensitive
to the approximation at the spin frequency where the pure quark
matter core forms and where we find the strongest effect of spin down.\\

In Figs.~\ref{Wt} and \ref{Wtscaled} we explore the importance of the
heating term in a little more detail. Fig.~\ref{Wt} shows the
numerical value of $W^\infty$ and its various contributions as a
function of time for a star with pairing gap $\Delta_{Q,0}=10$~MeV and
magnetic field $B=10^{11}$~G. The total entropy production,
$W^\infty$, is split into contributions corresponding to (integrals
of) the terms in Eq.~(\ref{dsdt}). $W^\infty_\textrm{Latent}$
corresponds to the latent heat, $W^\infty_\textrm{Hadronic bulk}$ to the
hadronic bulk term, $W^\infty_\textrm{Quark bulk}$ to the quark bulk
term and $W^\infty_\textrm{SC}$ to the surface and Coulomb term of
Eq.~(\ref{dsdt}). Leptons do not participate in the deconfinement
transition, so their entropy is assumed continuous across the phase
boundary. They are separated from the other particles in the form of
$W^\infty_\textrm{Lepton}$ in Fig.~\ref{Wt}.

 The bulk and lepton
contributions always act as heating terms while the surface and
Coulomb term changes sign and turns into a heating term with the onset
of hadronic superfluidity around age 4000 years with the hadronic bulk
term simultaneously falling away. The quark bulk term is very small
because for a pairing gap of 10 MeV the quark matter is strongly
superconducting at all times except the very early and the quark
entropy is therefore suppressed. This also implies, however, that
hadrons making the transition across the phase boundary essentially
release all their entropy, and so the latent heat, which changes sign
and becomes a heating term with the onset of quark superfluidity, is
therefore quite significant and actually comes to dominate around the
time when the hadronic bulk term falls away. The lepton term and the surface
and Coulomb term are relatively small but significant when the bulk
terms are eliminated through the onset of pairing. The peak in the
total around age
20000 years corresponds to the appearance of a pure quark matter phase
in the core and the ensuing rapid changes in structure, and it
contributes to produce the short interval of increasing temperature
seen in Fig.~\ref{Tt}.

Fig.~\ref{Wtscaled} shows the total entropy production scaled to the
total neutrino and photon luminosity, $W^\infty/(L^\infty _\gamma
+L^\infty _\nu)$, for the same set of parameters which gave the
cooling curves in Fig.~\ref{Tt}. The strong temperature dependence of
the neutrino and photon luminosities ensures that the scaled heating
term does not increase above unity where heating and cooling exactly
balance -- except at the peaks caused by the appearance of a pure
quark matter core. We also note that as in Fig.~\ref{Tt} the magnetic
field must be above $\sim 10^{9}$~G for $W^\infty$ to have any
discernible effects.

While the latent heat for a transition to a strongly superfluid core
is a heating term, the transition to a
nonsuperfluid core with higher entropy per baryon cools the
star. This is illustrated in Fig.~\ref{WtDelta} where we plot the
total heating term scaled to the combined neutrino and photon
luminosities at constant magnetic field for a range of quark pairing
gaps $\Delta_{Q,0}$. Here we see how the entropy production changes
from a heating term to a cooling term and may be quite dominant at
certain times depending on the choice of stellar parameters. This can
also be observed in Fig.~\ref{TtDelta} where we plot surface
temperature versus age with constant magnetic field for a range of
quark pairing gaps. The onset of superfluidity initially delays the
cooling and then accelerates it. When the pure quark matter core forms
there is a drop in temperature for very low pairing gaps and a jump
for high pairing gaps. In fact, however, the major
differences between the curves in Fig.~\ref{TtDelta} are due to the
differences in the thermal evolution of superfluid and nonsuperfluid
quark matter rather than spin down related effects. Old stars with a
nonsuperfluid quark matter core cool more slowly than those without a
quark matter core. 

A rich picture has now emerged from the coupling between the thermal
evolution and spin down compression with deconfinement. The spin down
may change the thermal evolution of hybrid stars by
heating and cooling them at various ages and by changing the structure
and chemical composition -- or it may be entirely inconsequential
depending on quark superfluidity and initial spin frequency. The
effects we found turned out to depend strongly on the inclusion of
quark pairing and the timing of the appearance of the pure quark
matter phase in the stellar core. These are subject to the
specific assumptions made concerning the equation of state, stellar
baryon number and pairing regime, so it is essential to stress that
the results discussed in the present section illustrate the general
point that spin down and thermal cooling may be interdependent, rather
than providing quantitatively reliable predictions.\\

\begin{figure}[!htb]
\centering
\plotone{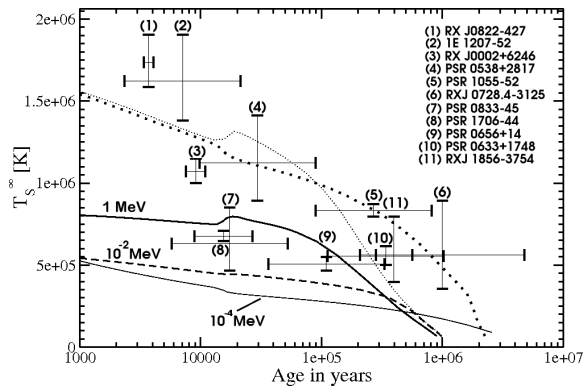}
\caption{Comparison between observational data and cooling
  calculations with spin down. The stars have initial spin frequency
  $\Omega_\mathrm{start}=6000 \mbox{ rad s}^{-1}$, magnetic field
  strength $B=10^{11}$~G and quark pairing gaps as indicated. The
  observational data can be found in \cite{Page:2004} and sources are
  identified by numbers right above or below their temperature
  error bars. The two dotted lines illustrate slow cooling with
  artificial suppression of neutrino emission by direct Urca and
  neutron pair breaking and formation. They have $\Delta_{Q,0}=10$~MeV
  (thick line) and 0.01 MeV (thin line). Including these mechanisms
  only cold sources can be explained. The jump in temperature at the
  formation of a pure quark matter core is smaller than the error bars
  on the observational data.}
\label{Ttwdata}
\end{figure}

Bearing this in mind we show in Fig.~\ref{Ttwdata} how calculations
for selected pairing gaps and magnetic field strengths measure up to
observational data on the thermal state of isolated neutron stars
(kindly supplied by Dany Page). This figure is shown partly as a
consistency check for our calculations and partly to compare the size
of the effects we have found with the accuracy of actual
measurements. We stress that the observational data are consistent
with static cooling models \citep{Page:2004} and that we do not
suggest that the effects of spin down are necessary to explain
them. Further the effects we have found occur after thermal relaxation
only for magnetic field strength below $\sim 10^{12}$~G. For those
sources in Fig.~\ref{Ttwdata} for which the field is known it is above
that value \citep{Pons:2007}.

Fig.~\ref{Ttwdata} explores a range of quark pairing gaps which in our
pairing scheme are density independent. In nature the pairing gap
would not span a range this broad, but it could be density dependent
and variations between stars in mass and density would then introduce
variations in their thermal evolution. Similarly stars with different
masses would have a different phase structure and acquire pure quark
matter cores at different rotational frequencies thus further
complicating the picture.

Because the hadronic direct Urca mechanism and the neutron pair
breaking and formation mechanism are active in the mixed phase our
models are consistent only with the cold sources and the effects of
spin down do not change this conclusion. That these mechanisms are the
cause of the generally low temperatures in our models can be seen if
they are artificially left out. The two dotted lines in
Fig.~\ref{Ttwdata} show -- for illustration only -- that temperatures
are much higher if these mechanisms are suppressed. Such a suppression
could occur for certain quark pairing schemes where pairing forces the
quark phase to become electrically neutral or positive rather than
negative, thereby reducing the proton content in the hadronic phase
below the threshold for direct Urca. A self-consistent inclusion of
such effects are beyond the scope of the present investigation. We
stress again that the inconsistency with hot sources seen in this
figure is subject to the choice of stellar and thermal models made for
the specific purpose of studying the effects of deconfinement during
spin down. Thus it should not be seen as indicating a general
breakdown in cooling theory.

The true signal of deconfinement we have found is in the transition
between cooling curves with and without pure quark matter cores, and
in the brief period of increasing or rapidly decreasing temperatures
thus caused for certain stellar parameters. As can be seen in
Fig.~\ref{Ttwdata} the change in surface temperature caused by
deconfinement is smaller than the error bars on the observational data
and it will be difficult to test observationally on the basis of data
relating only temperature and time.

\section{Discussion}\label{discuss}
Our intention with the present work was to explore a possible
connection between the spin down and thermal cooling of hybrid stars
with a deconfined quark matter core -- the appearance of which might be
expected to influence theoretical cooling tracks. Specifically we were
interested in the latent heat of the phase transition and the drastic
changes in structure and chemical composition resulting from the
increasing density with spin down. The general formalism worked out in
Sect.~\ref{thequilibrium} would also be relevant to models with no
phase transition, however, as the bulk terms in the spin down derived
entropy production might also be important in such cases. While
estimates of the importance of the spin down derived entropy
production did not give cause to expect strong signals of
deconfinement in the thermal evolution, numerical calculations
revealed clear effects of the changes caused by the appearance of a
pure quark matter core. This signature is related to changes in
radius, chemical composition, structure and under certain
circumstances the additional entropy production both in bulk and in
the form of latent heat.\\

Our numerical work was carried out using a sophisticated and reliable
code with respect to the stellar structure solving Hartles
perturbative equations for the structure of rotating compact stars
self-consistently. The thermal evolution, however, was treated in a
spherical isothermal approximation with a somewhat ad hoc approach to the
treatment of superfluid pairing in the quark matter phase. Still we believe our work is sufficiently detailed to demonstrate the
general point that important signals may be derived from the interplay
between spin down and cooling of compact stars, which could in the
long run help answer pressing questions about the state of matter at
high densities. The signals we have found do not dominate the general thermal
evolution of hybrid stars but they do complement the standard
picture. By correlating temperature with magnetic field strength and
spin frequency they may also help break the degeneracy between models relating
only temperature and time.\\

The signature of
deconfinement found here is below the present observational
sensitivity and not of sufficient strength to set apart the cooling
curves with temperature versus time for hybrid stars. The correlation
between temperature and magnetic field strength provides a possible
alternative. To test such a correlation it would be necessary to
obtain accurate temperatures for a number of stars of approximately
the same age for which the spin down could also be
detected by the emission of pulses in either radio or X-ray. It might
then be possible to test for features in the relation between
temperature and magnetic field -- or equally interesting, the rotation
frequency itself -- which as demonstrated here could result from a
strong phase transition if the initial rotation frequency was
sufficiently high to have spun out the high density phase. 

Contemplating these prospects it is important to consider that
alternative effects not treated here might determine the relation
between spin down and thermal evolution. Most pressingly the heat from
magnetic field decay was employed by \cite{Pons:2007} to explain the
observed correlation between temperature and magnetic field and it may
well drown out any other effects -- although as noted by the authors
the magnetic field decay itself has not yet been independently
demonstrated. Recent work by \cite{Aguilera:2007} and
\cite{Aguilera:2008} on the cooling of magnetized neutron stars in two
dimensions also highlighted the importance of the magnetic field
strength, geometry and possible decay for the thermal evolution of
neutron stars. These authors find strongly anisotropic surface
temperature distributions and possibly an inverted temperature
distribution with hot equatorial regions for middle aged stars. They
further show that the effects of magnetic field decay and Joule
heating can dominate the thermal evolution at strong and intermediate
field strengths -- even to the point that the effects of the direct
Urca process may be hidden by this heating term in magnetars.

The work of \cite{Reisenegger:1995} and \cite{Fernandez:2005} is also
most important. These authors consider the second term on the right
hand side of Eq.~(\ref{balance2}) which is shown to give rise to
so-called roto-chemical heating as the weak interactions required to
maintain chemical equilibrium are unable to keep pace with the
increasing density. \cite{Fernandez:2005} found that old millisecond
pulsars reach a quasi-equilibrium in which the photon luminosity is
determined entirely by the spin down power and remains much higher
than otherwise attainable. Their work considered only nucleonic
equations of state but similar results should clearly be expected for
hybrid stars just as part of the effects demonstrated here should be
expected to show up in nucleonic stars.\\

Given the possible importance of such alternatives and the
shortcomings of the present study discussed above, a more complete
treatment of the interplay between the spin down and thermal evolution
of neutron stars seems desirable before strong assertions can be made
concerning the specific shape of cooling tracks. As well as including
all possible energy sources and sinks this should be extended to
consider a wider range of stellar masses, equations of state, phase
transitions and pairing regimes. A two dimensional code could further
treat the influence of the magnetic field on heat flows in the inner
crust and the effects of nonspherical heat flows for rotationally
perturbed stars -- aspects which are all
independently well described in the literature.

It should further be noted that the rotation frequency at which
the pure quark matter core appears depends strongly on the stellar
baryon number and the equation of state, and it may well be lower than
discussed here thus changing the timing of spin down related
effects. The strongest signature of deconfinement found here depends
entirely on the formation of a pure quark matter core where none
existed previously. If, on the other hand, a pure quark matter phase
is present in the core of stars at even the highest rotation
frequencies -- as is the case for other equations of state -- we
would not expect equally strong signals from deconfinement with spin
down compression.

It is possible that circumstances not considered here could
provide a more pronounced link between rotation and cooling. If for
instance the magnetic field varies over time -- and as discussed by
\cite{Geppert:2006} this may well be the case -- the timing of any
strengthening or decay of the magnetic field may provide entirely
different spin down histories. Alternatively the temperature of stars
being spun up in accreting binaries should to some extent be
determined by the changing chemical composition and this could provide
an alternative approach. 

Additional energy release would also be expected if the
phase transition cannot maintain thermodynamic or hydrostatic
equilibrium as was assumed here. As the most extreme example of this,
we have entirely ignored the instabilities and corequakes which might
accompany the appearance of a pure quark matter phase in a star far
from equilibrium \citep{Zdunik:2006}. Such events could release large
amounts of energy and significantly change the thermal
evolution by reheating the star.\\
     
While many essential issues thus remain poorly explored we hope to
have demonstrated the possible usefulness of considering the
connection between the spin down and thermal evolution of neutron
stars. Although the results of such work may
not be immediately applicable to observational tests, we believe they
could prove most valuable in the long run.\\

We wish to acknowledge helpful conversation with Sanjay Reddy on the
hadronic pair breaking and formation process and to thank Dany Page
for the use of his compilation of observational data. We are also
grateful to Dima Yakovlev and Oleg Gnedin for providing us with an
earlier version of their cooling code, which we took advantage of when
developing the code used for this study.\\
M. Stejner wishes to thank San Diego State University for its
hospitality during part of this work.\\
The research of F. Weber is supported by the National Science
Foundation under Grant PHY-0457329, and by the Research Corporation.\\
J. Madsen is supported by the Danish Natural Science Research Council.

\appendix \section{Appendix A, Derivation of the Energy Balance in Mixed
Phases}\label{appendix} 

To derive Eq.~(\ref{balance2}) from
Eq.~(\ref{balance}) we proceed analogously to \cite{Thorne:1966} and
\cite{Weber:1999} but work in terms of baryon number $a$ instead of
radial coordinate. We first define the nuclear energy generation rate
per baryon as the rate, $q$, at which rest mass is converted to
internal energy as measured by an observer in the shell and at rest
with respect to the shells reference frame
\begin{equation}\label{first}
q=-\frac{\mathrm{d}\bar{m}_\mathrm{B}}{\mathrm{d}(\mbox{proper
    time})}=-\frac{\mathrm{d}\bar{m}_\mathrm{B}}{e^\Phi \mathrm{d}t}\; ,
\end{equation}
where $\bar{m}_\mathrm{B}$ is the average rest mass per baryon and we
work in units with. The
first term on the right hand side of Eq.~(\ref{balance}) is the amount
of rest mass converted to internal energy in the entire shell which is
then $q\delta a e^\Phi \mathrm{d}t$. 

Next we note that by expressing the shells volume in terms of its
particle density, $V=(V/\delta a)\delta a=\rho/\delta a$, the second term in
Eq.~(\ref{balance}) can be written
\begin{equation}\label{second}
\mathrm{d}W=-P\mathrm{d}\left(\frac{1}{\rho}\delta a\right) + \alpha \mathrm{d}\mathbb{S}+\mathrm{d}E_\mathrm{C}
\end{equation}  

The third term represents the difference between the rates at which
energy enters and leaves the shell by radiative or conductive means as
measured by an observer in the shell and at rest with respect to the
shells reference frame, which may be written
\begin{eqnarray} \label{third}
L_\mathrm{tot}(a+\delta a)e^{2(\Phi(a+\delta
  a)-\Phi(a))}-L_\mathrm{tot}(a)&=&L_\mathrm{tot}(a+\delta
a)\left(1+2\frac{\mathrm{d}\Phi}{\mathrm{d}a}\delta
  a\right)-L_\mathrm{tot}(a)\nonumber\\
&=&\left(\frac{\mathrm{d}L_\mathrm{tot}}{\mathrm{d}
  a}+2L_\mathrm{tot}\frac{\mathrm{d} \Phi}{\mathrm{d} a}\right)\delta a\nonumber\\
&=&\frac{\mathrm{d}}{\mathrm{d}
  a}\left(L_\mathrm{tot}e^{2\Phi}\right)e^{-2\Phi}\delta a 
\end{eqnarray}
and the energy change during a coordinate time interval, d$t$, is this
times the proper time interval $e^\Phi \mathrm{d}t$. The two factors
of $e^{\Phi(a+\delta a)-\Phi(a)}$ account for redshift and time
dilation, respectively, as the energy crosses from the inner to the outer
edge of the shell, and we expanded the exponential to order $
\mathcal{O}(\delta a)$ as 
\begin{equation}
e^{2(\Phi(a+\delta a)-\Phi(a))}\approx 1+2[\Phi(a+\delta a)-\Phi(a)]=
1+2\frac{\mathrm{d}\Phi}{\mathrm{d} a}\delta a.
\end{equation}

Using Eqs.~(\ref{first}), (\ref{second}) and (\ref{third}) in
Eq.~(\ref{balance}) and expressing the change in internal energy in
terms of the density of internal energy, $\mathcal{E}_\mathrm{int}$,
local energy balance can then be written as
\begin{equation}
\mathrm{d}\left(\frac{\mathcal{E}_\mathrm{int}}{\rho}\delta
a\right)=qe^\Phi\delta a \mathrm{d}t -
P\mathrm{d}\left(\frac{1}{\rho}\delta a\right) + \alpha
\mathrm{d}\mathbb{S} + \mathrm{d}E_\mathrm{C} - \left[\frac{\mathrm{d}}{\mathrm{d}
  a}\left(L_\mathrm{tot}e^{2\Phi}\right)\right] e^{-\Phi}\delta a \mathrm{d}t\;.
\end{equation}
Rearranging and noting that $\delta a$ is constant in time we then get the
gradient of $L_\mathrm{tot}$
\begin{eqnarray}\label{cool1}
\frac{\mathrm{d}}{\mathrm{d}
  a}\left(L_\mathrm{tot}e^{2\Phi}\right)=e^{2\Phi}\bigg\{q+\bigg[\frac{Pe^{-\Phi}}{\rho^2}\frac{\mathrm{d}\rho}{\mathrm{d}t}&+&\frac{\alpha
  e^{-\Phi}}{\delta
  a}\frac{\mathrm{d}\mathbb{S}}{\mathrm{d}t}\nonumber\\ &+&\frac{e^{-\phi}}{\delta
  a}\frac{\mathrm{d}E_\mathrm{C}}{\mathrm{d} t}-e^{-\Phi}\frac{\mathrm{d}}{\mathrm{d}t}\left(\frac{\mathcal{E}_\mathrm{int}}{\rho}\right)\bigg]_{a=\mathrm{constant}}\bigg\}
\end{eqnarray}

Eq.~(\ref{cool1}) is the relation we sought giving the luminosity
gradient for stars with variable structure and including the
contributions from surface and Coulomb terms in the energy density. It
may be considerably simplified however by noting that we can write the
internal energy density in terms of the total energy density as
$\mathcal{E}_\mathrm{int}/\rho=\epsilon/\rho-\bar{m}_\mathrm{B}$ in
which case from the definition of $q$
\begin{equation}
q-e^{-\Phi}\frac{\mathrm{d}}{\mathrm{d}t}\left(\frac{\mathcal{E}_\mathrm{int}}{\rho}\right)=
-\frac{e^{-\Phi}}{\rho}\left[\frac{\mathrm{d}\epsilon}{\mathrm{d}t}-\frac{\epsilon}{\rho}\frac{\mathrm{d}\rho}{\mathrm{d}t}\right]\; .
\end{equation}
Inserting this in Eq.~(\ref{cool1}) and using the first law as written
in Eq.~(\ref{firstlaw}) finally gives the expression
\begin{eqnarray}\label{cool2}
\frac{\mathrm{d}}{\mathrm{d}
  a}\left(L_\mathrm{tot}e^{2\Phi}\right)&=&-\frac{e^{\Phi}}{\rho}\left[\frac{\mathrm{d}\epsilon}{\mathrm{d}t}-\frac{\epsilon+P}{\rho}\frac{\mathrm{d}\rho}{\mathrm{d}t} 
-\rho\frac{\alpha}{\delta a}\frac{\mathrm{d}\mathbb{S}}{\mathrm{d}t}
-\rho\frac{1}{\delta a}\frac{\mathrm{d}E_\mathrm{C}}{\mathrm{d}
  t}\right]_{a=\mathrm{constant}}\\
&=&-e^\Phi\left[T\frac{\mathrm{d}s}{\mathrm{d}t}+\sum_k\mu_k\frac{\mathrm{d}Y_k}{\mathrm{d}t}\right]_{a=\mathrm{constant}}
  \; .
\end{eqnarray}

In this discussion we have used baryon number as the
independent variable to emphasize the importance of a Lagrangian
description for stars with variable structure. This is of course
not required for all applications and for comparison with other
discussions we note that our expressions can be converted to use radial
coordinate as the independent coordinate through the standard relation
\begin{equation}
\mathrm{d} a=4\pi r^2\rho\left(1-\frac{2m}{r}\right)^{-\frac{1}{2}}\mathrm{d}r
\end{equation} 


\begin{thebibliography}{69}
\expandafter\ifx\csname natexlab\endcsname\relax\def\natexlab#1{#1}\fi

\bibitem[{{Aguilera} {et~al.}(2008){Aguilera}, {Pons}, \&
  {Miralles}}]{Aguilera:2007}
{Aguilera}, D.~N., {Pons}, J.~A., \& {Miralles}, J.~A. 2008,
  \aap, 486, 255

\bibitem[{{Aguilera} {et~al.}(2008){Aguilera}, {Pons}, \&
  {Miralles}}]{Aguilera:2008}
{Aguilera}, D.~N., {Pons}, J.~A., \& {Miralles}, J.~A. 2008, \apjl, 673, L167

\bibitem[{{Alford}(2001)}]{Alford:2001}
{Alford}, M. 2001, Annual Review of Nuclear and Particle Science, 51, 131

\bibitem[{{Alford} {et~al.}(1999){Alford}, {Rajagopal}, \&
  {Wilczek}}]{Alford:1999}
{Alford}, M., {Rajagopal}, K., \& {Wilczek}, F. 1999, Nuclear Physics B, 537,
  443

\bibitem[{{Alford} {et~al.}(2008){Alford}, {Rajagopal}, {Schaefer}, \&
  {Schmitt}}]{Alford:2007a}
{Alford}, M.~G., {Rajagopal}, K., {Schaefer}, T., \& {Schmitt}, A. 2008,
  Reviews of Modern Physics, 80, 1455

\bibitem[{{Andersson} {et~al.}(2005){Andersson}, {Comer}, \&
  {Glampedakis}}]{Andersson:2005}
{Andersson}, N., {Comer}, G.~L., \& {Glampedakis}, K. 2005, Nuclear Physics A,
  763, 212

\bibitem[{{Baldo} {et~al.}(2003){Baldo}, {Burgio}, \& {Schulze}}]{Baldo:2003}
{Baldo}, M., {Burgio}, G.~F., \& {Schulze}, H.~. 2003, [ArXiv:astro-ph/0312446]

\bibitem[{{Baym}(2007)}]{Baym:2006}
{Baym}, G. 2007, AIP Conference Proceedings, 892, 8 

\bibitem[{{Berger} \& {Jaffe}(1987)}]{Berger:1987}
{Berger}, M.~S. \& {Jaffe}, R.~L. 1987, \prc, 35, 213

\bibitem[{{Blaschke} {et~al.}(2000){Blaschke}, {Kl{\"a}hn}, \&
  {Voskresensky}}]{Blaschke:2000}
{Blaschke}, D., {Kl{\"a}hn}, T., \& {Voskresensky}, D.~N. 2000, \apj, 533, 406

\bibitem[{{Drago} {et~al.}(2008){Drago}, {Pagliara}, \& {Parenti}}]{Drago:2008}
{Drago}, A., {Pagliara}, G., \& {Parenti}, I. 2008, \apjl, 678, L117

\bibitem[{{Endo} {et~al.}(2006){Endo}, {Maruyama}, {Chiba}, \&
  {Tatsumi}}]{Endo:2006}
{Endo}, T., {Maruyama}, T., {Chiba}, S., \& {Tatsumi}, T. 2006, Progress of
  Theoretical Physics, 115, 337

\bibitem[{{Fern{\'a}ndez} \& {Reisenegger}(2005)}]{Fernandez:2005}
{Fern{\'a}ndez}, R. \& {Reisenegger}, A. 2005, \apj, 625, 291

\bibitem[{{Freire}(2008)}]{Freire:2007a}
{Freire}, P.~C.~C. 2008, AIP Conference Proceedings, 983, 459

\bibitem[{{Freire} {et~al.}(2008{\natexlab{a}}){Freire}, {Ransom}, {B{\'e}gin},
  {Stairs}, {Hessels}, {Frey}, \& {Camilo}}]{Freire:2008a}
{Freire}, P.~C.~C., {Ransom}, S.~M., {B{\'e}gin}, S., {et~al.}
  2008{\natexlab{a}}, \apj, 675, 670

\bibitem[{{Freire} {et~al.}(2008{\natexlab{b}}){Freire}, {Ransom}, {B{\'e}gin},
  {Stairs}, {Hessels}, {Frey}, \& {Camilo}}]{Freire:2008}
{Freire}, P.~C.~C., {Ransom}, S.~M., {B{\'e}gin}, S., {et~al.}
  2008{\natexlab{b}}, in American Institute of Physics Conference Series, Vol.
  983, 40 Years of Pulsars: Millisecond Pulsars, Magnetars and
  More, ed. C. Bassa, Z. Wang, A. Cumming, \& V. M. Kaspi (Melville: AIP), 604

\bibitem[{{Geppert}(2006)}]{Geppert:2006}
{Geppert}, U. 2006, [ArXiv:astro-ph/0611708]

\bibitem[{{Geppert} {et~al.}(2004){Geppert}, {K{\"u}ker}, \&
  {Page}}]{Geppert:2004}
{Geppert}, U., {K{\"u}ker}, M., \& {Page}, D. 2004, \aap, 426, 267

\bibitem[{{Glendenning}(1990)}]{Glendenning:1990}
{Glendenning}, N.~K. 1990, Nuclear Physics A, 512, 737

\bibitem[{{Glendenning}(1992)}]{Glendenning:1992}
{Glendenning}, N.~K. 1992, \prd, 46, 1274

\bibitem[{{Glendenning}(2000)}]{Glendenning:2000}
{Glendenning}, N.~K. 2000, {Compact stars : nuclear physics, particle physics,
  and general relativity} (Berlin: Springer)

\bibitem[{{Glendenning} \& {Weber}(2001)}]{Glendenning:2001a}
{Glendenning}, N.~K. \& {Weber}, F. 2001, \apjl, 559, L119

\bibitem[{{Gudmundsson} {et~al.}(1983){Gudmundsson}, {Pethick}, \&
  {Epstein}}]{Gudmundsson:1983}
{Gudmundsson}, E.~H., {Pethick}, C.~J., \& {Epstein}, R.~I. 1983, \apj, 272,
  286

\bibitem[{{Hartle}(1967)}]{Hartle:1967}
{Hartle}, J.~B. 1967, \apj, 150, 1005

\bibitem[{{Hartle} \& {Thorne}(1968)}]{Hartle:1968}
{Hartle}, J.~B. \& {Thorne}, K.~S. 1968, \apj, 153, 807

\bibitem[{{Heiselberg} {et~al.}(1993){Heiselberg}, {Pethick}, \&
  {Staubo}}]{Heiselberg:1993}
{Heiselberg}, H., {Pethick}, C.~J., \& {Staubo}, E.~F. 1993, Physical Review
  Letters, 70, 1355

\bibitem[{{Jaikumar} {et~al.}(2002){Jaikumar}, {Prakash}, \&
  {Sch{\"a}fer}}]{Jaikumar:2002}
{Jaikumar}, P., {Prakash}, M., \& {Sch{\"a}fer}, T. 2002, \prd, 66, 063003

\bibitem[{{Jaikumar} \& {Prakash}(2001)}]{Jaikumar:2001}
{Jaikumar}, P. \& {Prakash}, S. 2001, Physics Letters B, 516, 345

\bibitem[{{Kaaret} {et~al.}(2007){Kaaret}, {Prieskorn}, {Zand}, {Brandt},
  {Lund}, {Mereghetti}, {G{\"o}tz}, {Kuulkers}, \& {Tomsick}}]{Kaaret:2007}
{Kaaret}, P., {Prieskorn}, Z., {Zand}, J.~J.~M.~i., {et~al.} 2007, \apjl, 657,
  L97

\bibitem[{{Kaminker} {et~al.}(2001){Kaminker}, {Haensel}, \&
  {Yakovlev}}]{Kaminker:2001}
{Kaminker}, A.~D., {Haensel}, P., \& {Yakovlev}, D.~G. 2001, \aap, 373, L17

\bibitem[{{Landau} \& {Lifshitz}(1980)}]{Landau:1980}
{Landau}, L.~D. \& {Lifshitz}, E.~M. 1980, {Statistical physics. Pt.1, Pt.2},
  3rd edn. (Oxford: Pergamon Press)

\bibitem[{{Lattimer} \& {Prakash}(2007)}]{Lattimer:2007}
{Lattimer}, J.~M. \& {Prakash}, M. 2007, \physrep, 442, 109

\bibitem[{{Lattimer} {et~al.}(1991){Lattimer}, {Prakash}, {Pethick}, \&
  {Haensel}}]{Lattimer:1991}
{Lattimer}, J.~M., {Prakash}, M., {Pethick}, C.~J., \& {Haensel}, P. 1991,
  Physical Review Letters, 66, 2701

\bibitem[{{Leinson} \& {P{\'e}rez}(2006)}]{Leinson:2006}
{Leinson}, L.~B. \& {P{\'e}rez}, A. 2006, Physics Letters B, 638, 114

\bibitem[{{Lombardo} \& {Schulze}(2001)}]{Lombardo:2001}
{Lombardo}, U. \& {Schulze}, H.-J. 2001, LNP, 578, 30

\bibitem[{{Miao} \& {Xiao-Ping}(2007)}]{Miao:2007}
{Miao}, K. \& {Xiao-Ping}, Z. 2007, \mnras, 375, 1503

\bibitem[{{Miao} {et~al.}(2007){Miao}, {Xiao-Ping}, \& {Na-Na}}]{Miao:2007a}
{Miao}, K., {Xiao-Ping}, Z., \& {Na-Na}, P. 2007, [ArXiv:astro-ph/0708.0900]

\bibitem[{{Miralles} \& {van Riper}(1996)}]{Miralles:1996}
{Miralles}, J.~A. \& {van Riper}, K.~A. 1996, \apjs, 105, 407

\bibitem[{{Miralles} {et~al.}(1993){Miralles}, {van Riper}, \&
  {Lattimer}}]{Miralles:1993}
{Miralles}, J.~A., {van Riper}, K.~A., \& {Lattimer}, J.~M. 1993, \apj, 407,
  687

\bibitem[{{M{\"u}hlschlegel}(1959)}]{Muhlschlegel:1959}
{M{\"u}hlschlegel}, B. 1959, Zeitschrift f{\"u}r Physik, 155, 313

\bibitem[{{Niebergal} {et~al.}(2007){Niebergal}, {Ouyed}, \&
  {Leahy}}]{Niebergal:2007}
{Niebergal}, B., {Ouyed}, R., \& {Leahy}, D. 2007, \aap, 476, L5

\bibitem[{{\"O}zel(2006)}]{Ozel:2006}
{\"O}zel, F. 2006, Nature, 441, 1115

\bibitem[{{Page} {et~al.}(2006){Page}, {Geppert}, \& {Weber}}]{Page:2006a}
{Page}, D., {Geppert}, U., \& {Weber}, F. 2006, Nuclear Physics A, 777, 497

\bibitem[{{Page} {et~al.}(2004){Page}, {Lattimer}, {Prakash}, \&
  {Steiner}}]{Page:2004}
{Page}, D., {Lattimer}, J.~M., {Prakash}, M., \& {Steiner}, A.~W. 2004, \apjs,
  155, 623

\bibitem[{{Page} \& {Reddy}(2006)}]{Page:2006}
{Page}, D. \& {Reddy}, S. 2006, Annual Review of Nuclear and Particle Science,
  56, 327

\bibitem[{{Pons} {et~al.}(2007){Pons}, {Link}, {Miralles}, \&
  {Geppert}}]{Pons:2007}
{Pons}, J.~A., {Link}, B., {Miralles}, J.~A., \& {Geppert}, U. 2007, Physical
  Review Letters, 98, 071101

\bibitem[{{Potekhin} {et~al.}(1997){Potekhin}, {Chabrier}, \&
  {Yakovlev}}]{Potekhin:1997}
{Potekhin}, A.~Y., {Chabrier}, G., \& {Yakovlev}, D.~G. 1997, \aap, 323, 415

\bibitem[{{Potekhin} \& {Yakovlev}(2001)}]{Potekhin:2001}
{Potekhin}, A.~Y. \& {Yakovlev}, D.~G. 2001, \aap, 374, 213

\bibitem[{{Reisenegger}(1995)}]{Reisenegger:1995}
{Reisenegger}, A. 1995, \apj, 442, 749

\bibitem[{{Schaab} {et~al.}(1999){Schaab}, {Sedrakian}, {Weber}, \&
  {Weigel}}]{Schaab:1999}
{Schaab}, C., {Sedrakian}, A., {Weber}, F., \& {Weigel}, M.~K. 1999, \aap, 346,
  465

\bibitem[{{Schaab} {et~al.}(1996){Schaab}, {Weber}, {Weigel}, \&
  {Glendenning}}]{Schaab:1996}
{Schaab}, C., {Weber}, F., {Weigel}, M.~K., \& {Glendenning}, N.~K. 1996,
  Nuclear Physics A, 605, 531

\bibitem[{{Schaab} \& {Weigel}(1998)}]{Schaab:1998a}
{Schaab}, C. \& {Weigel}, M.~K. 1998, \aap, 336, L13

\bibitem[{{Schaffner-Bielich}(2007)}]{Schaffner-Bielich:2007}
{Schaffner-Bielich}, J. 2007, [ArXiv:astro-ph/0703113]

\bibitem[{{Schmitt} {et~al.}(2002){Schmitt}, {Wang}, \&
  {Rischke}}]{Schmitt:2002}
{Schmitt}, A., {Wang}, Q., \& {Rischke}, D.~H. 2002, \prd, 66, 114010

\bibitem[{{Schulze} {et~al.}(2006){Schulze}, {Polls}, {Ramos}, \&
  {Vida{\~n}a}}]{Schulze:2006}
{Schulze}, H.-J., {Polls}, A., {Ramos}, A., \& {Vida{\~n}a}, I. 2006, \prc, 73,
  058801

\bibitem[{{Steiner} \& {Reddy}(2009)}]{Steiner:2008}
{Steiner}, A.~W. \& {Reddy}, S. 2009, \prc, 79, 015802

\bibitem[{{Steiner} {et~al.}(2002){Steiner}, {Reddy}, \&
  {Prakash}}]{Steiner:2002}
{Steiner}, A.~W., {Reddy}, S., \& {Prakash}, M. 2002, \prd, 66, 094007

\bibitem[{{Thorne}(1966)}]{Thorne:1966}
{Thorne}, K.~S. 1966, in Proc. Int School of Physics
  Enrico Fermi XXXV, ed. L.~{Gratton}, (Academic: New York), 166

\bibitem[{{Van Riper}(1991)}]{van-Riper:1991}
{Van Riper}, K.~A. 1991, \apjs, 75, 449

\bibitem[{{Voskresensky} {et~al.}(2003){Voskresensky}, {Yasuhira}, \&
  {Tatsumi}}]{Voskresensky:2003}
{Voskresensky}, D.~N., {Yasuhira}, M., \& {Tatsumi}, T. 2003, Nuclear Physics
  A, 723, 291

\bibitem[{{Weber}(1999)}]{Weber:1999}
{Weber}, F. 1999, {Pulsars as astrophysical laboratories for nuclear and
  particle physics} (IOP Publishing)

\bibitem[{{Weber}(2005)}]{Weber:2005}
{Weber}, F. 2005, Progress in Particle and Nuclear Physics, 54, 193

\bibitem[{{Weber} \& {Glendenning}(1992)}]{Weber:1992}
{Weber}, F. \& {Glendenning}, N.~K. 1992, \apj, 390, 541

\bibitem[{{Weber} {et~al.}(1991){Weber}, {Glendenning}, \&
  {Weigel}}]{Weber:1991}
{Weber}, F., {Glendenning}, N.~K., \& {Weigel}, M.~K. 1991, \apj, 373, 579

\bibitem[{{Xiaoping} {et~al.}(2008){Xiaoping}, {Li}, {Xia}, \&
  {Miao}}]{Xiaoping:2008}
{Xiaoping}, Z., {Li}, Z., {Xia}, Z., \& {Miao}, K. 2008, ArXiv e-prints, 0808.1587

\bibitem[{{Yakovlev} {et~al.}(2001){Yakovlev}, {Kaminker}, {Gnedin}, \&
  {Haensel}}]{Yakovlev:2001}
{Yakovlev}, D.~G., {Kaminker}, A.~D., {Gnedin}, O.~Y., \& {Haensel}, P. 2001,
  \physrep, 354, 1

\bibitem[{{Yakovlev} {et~al.}(1999){Yakovlev}, {Levenfish}, \&
  {Shibanov}}]{Yakovlev:1999}
{Yakovlev}, D.~G., {Levenfish}, K.~P., \& {Shibanov}, Y.~A. 1999, Soviet
  Physics Uspekhi, 42, 737

\bibitem[{{Yakovlev} \& {Pethick}(2004)}]{Yakovlev:2004a}
{Yakovlev}, D.~G. \& {Pethick}, C.~J. 2004, \araa, 42, 169

\bibitem[{{Zdunik} {et~al.}(2006){Zdunik}, {Bejger}, {Haensel}, \&
  {Gourgoulhon}}]{Zdunik:2006}
{Zdunik}, J.~L., {Bejger}, M., {Haensel}, P., \& {Gourgoulhon}, E. 2006, \aap,
  450, 747

\end{thebibliography}

\end{document}